\documentclass[twocolumn,reprint,aps,prl]{revtex4-1}
\usepackage[utf8]{inputenc}
\usepackage{color}
\usepackage{amsmath}
\usepackage{amssymb}
\usepackage{amsbsy}
\usepackage{bbold}
\usepackage{graphicx}
\usepackage{xfrac}
\usepackage[]{todonotes}
\usepackage[unicode=true,
 bookmarks=false,
 breaklinks=false,pdfborder={0 0 1},backref=false,colorlinks=true]
 {hyperref}
\hypersetup{
 linkcolor=blue,citecolor=blue,urlcolor=blue,linkcolor=blue,citecolor=blue,urlcolor=blue}

\makeatletter

\usepackage{braket}

\definecolor{mygreen}{rgb}{0,0.5,0}
\definecolor{myblue}{rgb}{0,0,0.75}
\definecolor{mymagenta}{cmyk}{0,1,0,0.12}

\makeatother

\begin{document}

\title{Subradiant Bell states in distant atomic arrays}

\author{P.-O.~Guimond}
\affiliation{Center for Quantum Physics,
University of Innsbruck, Innsbruck A-6020, Austria}
\affiliation{Institute for Quantum Optics and Quantum Information, Austrian Academy of Sciences, Innsbruck A-6020,
      Austria}
     
\author{A.~Grankin}
\affiliation{Center for Quantum Physics,
University of Innsbruck, Innsbruck A-6020, Austria}
\affiliation{Institute for Quantum Optics and Quantum Information, Austrian Academy of Sciences, Innsbruck A-6020,
      Austria}

\author{D.\,V.~Vasilyev}
\affiliation{Center for Quantum Physics,
University of Innsbruck, Innsbruck A-6020, Austria}
\affiliation{Institute for Quantum Optics and Quantum Information, Austrian Academy of Sciences, Innsbruck A-6020,
      Austria}
      
\author{B.~Vermersch}
\affiliation{Center for Quantum Physics,
University of Innsbruck, Innsbruck A-6020, Austria}
\affiliation{Institute for Quantum Optics and Quantum Information, Austrian Academy of Sciences, Innsbruck A-6020,
      Austria}

\author{P.~Zoller}
\affiliation{Center for Quantum Physics,
University of Innsbruck, Innsbruck A-6020, Austria}
\affiliation{Institute for Quantum Optics and Quantum Information, Austrian Academy of Sciences, Innsbruck A-6020,
      Austria}

\begin{abstract}
We study collective `free-space' radiation properties of two distant single-layer arrays of quantum emitters as two-level atoms. We show that this system can support a long-lived Bell superposition state of atomic excitations exhibiting strong subradiance, which corresponds to a non-local excitation of the two arrays. We describe the preparation of these states and their application in quantum information as resource of non-local entanglement, including deterministic quantum state transfer with high fidelity between the arrays representing quantum memories. We discuss experimental realizations using cold atoms in optical trap arrays with subwavelength spacing, and analyze the role of imperfections. 
\end{abstract}
\maketitle

\textit{Introduction. ---}
Recent advances in preparing regular arrays of atoms with optical traps \cite{Lester2015,Xia2015,Endres2016,Barredo2016} offer new opportunities to engineer strong collective coupling between atoms and light, with applications in quantum information science. In particular, a single layer of atoms loaded into a regular 2D array with sub-wavelength spacing has been proposed as an atomic mirror with high reflectivity \cite{GarciaDeAbajo2007,PhysRevLett.111.147401,Bettles2016,PhysRevLett.119.053901, Shahmoon2017},  as quantum memory with efficient storage and retrieval \cite{Manzoni2017}, and to implement topological quantum optics \cite{Perczel2017,Bettles2017}; in addition, emission of single photons from bilayer atomic arrays can be engineered to be highly directional in free-space \cite{Grankin2018}.  Moreover, single-layered atomic arrays have been shown to support \emph{subradiant} collective excitations \cite{Facchinetti2016,Plankensteiner2017,Asenjo-Garcia2017}, which consist of excited superposition states of atoms decaying much slower than a single isolated excited atom, due to interference in spontaneous emission \cite{Dicke1953,Scully2015,Guerin2016,Sutherland2016,Jen2016,Solano2017,Moreno-Cardoner2019}.

\begin{figure}
\includegraphics[width=\columnwidth]{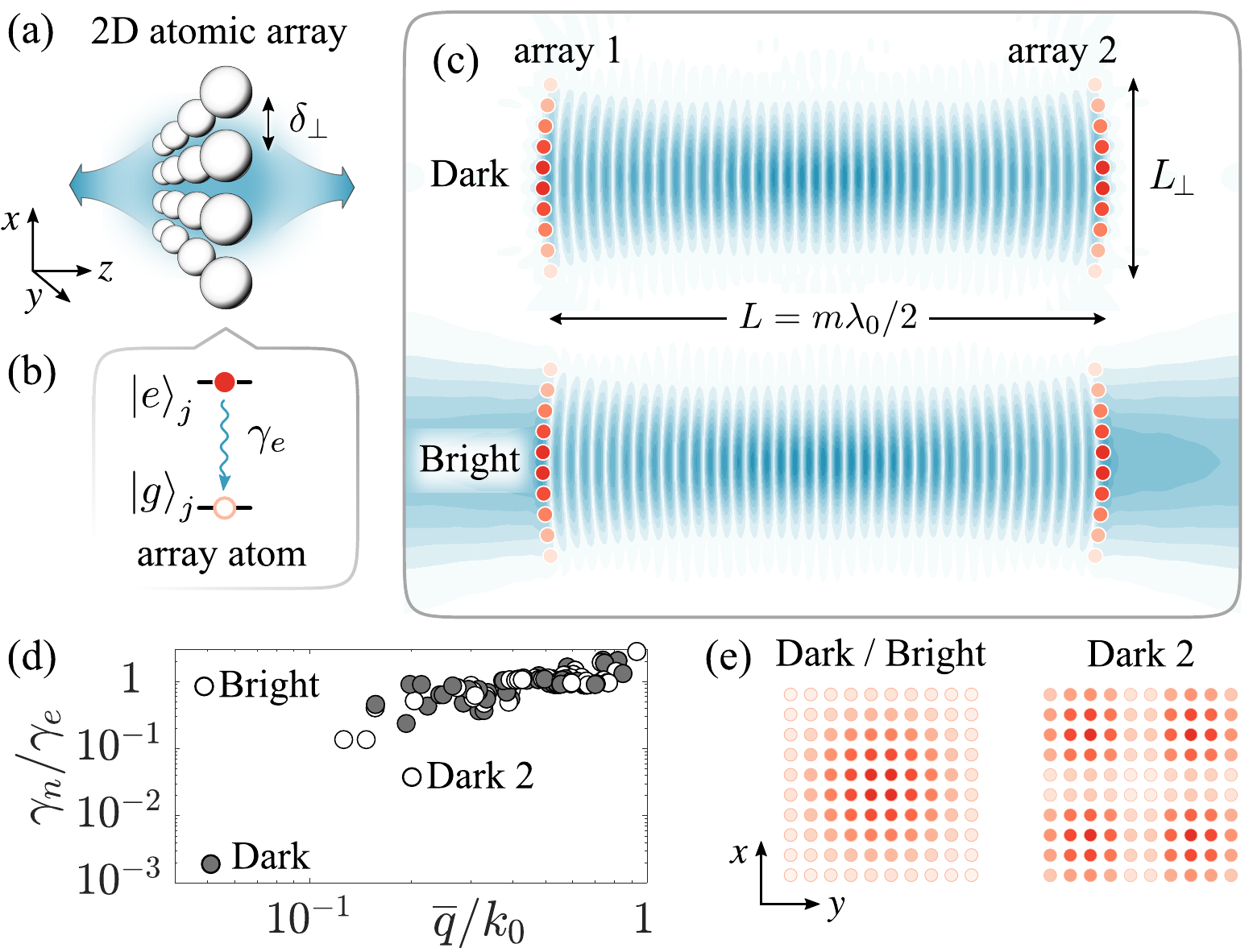} 
	\caption{\textit{`Dark' and `bright' states in two distant atomic arrays.} (a)~Sketch of a \textit{single} 2D atomic array, with light emitted perpendicular to atomic plane (corresponding to a `bright', i.e. radiating state). (b)~Two-level scheme. (c)~Setup with two distant atomic arrays: we plot the electric field profile $|{\boldsymbol \psi}({\boldsymbol r})|$ of photonic modes (blue) associated with the `dark' and `bright' states as excitations in the two arrays (red). (d)~Decay rates  $\gamma_n$ as imaginary part of eigenenergies of the non-hermitian effective Hamiltonian $\mathcal H$ [Eq.~\eqref{eq:mathbbH}], in units of the single atom decay rate $\gamma_e$, ordered according to their quasi-momentum $\overline q$ (see text). The white (black) color denotes even (odd) parity. A pair of `dark' and `bright' states are identified as the left-most dots. (e)~Atomic wavefunction amplitudes in each array $|(v_n)_{{\boldsymbol j}_\perp}|$ associated with the dark and bright state. In (c-e) $\delta_\perp=0.75\lambda_0$, {$N_\perp=10$}, $L=20\lambda_0$ (see text). 
	\label{fig:setup} }
\end{figure}

Here we show that the composite quantum system consisting of two distant single-layered arrays of atoms [cf.~Figs.~\ref{fig:setup}(a-c)] can support an atomic Bell superposition state exhibiting strong subradiance. Remarkably, this non-radiating `dark' state is a \emph{non-local} entangled state, i.e.~a superposition state of a collective excitation living in the first \textit{or} second array, where the two arrays can be separated by a distance $L$ much larger than the transverse size $L_\perp$ of each individual array. This phenomenon relies on two ingredients. First, spontaneous emission from a collective atomic excitation in a single layer can be directional, with a proper phasing of the atomic dipoles, corresponding to light emission in both directions perpendicular to the atomic array, as in Fig.~\ref{fig:setup}(a) \cite{GarciaDeAbajo2007}. Second, radiation from two distant atomic arrays can -- provided the separation length $L$ is commensurate with half the optical wavelength [upper panel in Fig.~\ref{fig:setup}(c)] -- lead to destructive interference of light emitted to the left and to the right of the two arrays, corresponding to a subradiant state, i.e.~this `dark' state will show strongly suppressed radiative loss to the outside world.  
In contrast,  the lower panel in Fig.~\ref{fig:setup}(c) displays a `bright' (i.e.,~radiating) state due to constructive interference. 

Below we will show that these non-local subradiant atomic superposition states can be prepared naturally in setups involving two -- or more -- atomic arrays, and provide a source of entanglement shared between the two atomic arrays, with applications for quantum networking \cite{Northup2014}. In particular, quantum information can be exchanged between the arrays representing `local' quantum memories, in a \emph{coherent} and \emph{deterministic} process, with dark states acting as mediators.

\textit{Quantum optical model. ---}
Our setup consists of two 2D arrays of {$N=N_\perp\times N_\perp$} atomic emitters with lattice spacing $\delta_\perp$ and size $L_\perp\equiv N_\perp \delta_\perp$, separated by a distance $L$ along $z$. Each atom has a ground and an excited state, $\ket{g}_{\boldsymbol j}$ and $\ket{e}_{\boldsymbol j}$, and is coupled to free-space modes of the radiation field via a dipole transition with frequency $\omega_0=ck_0=2\pi c/\lambda_0$. Here the multi-index ${\boldsymbol j}=({\boldsymbol j}_\perp,j_z)$, where $j_z=1,2$ labels the arrays, while ${\boldsymbol j}_\perp=(j_x,j_y)$ label the atoms within each array, with $1\leq j_x,j_y\leq N_\perp$. Atomic positions are denoted by ${\boldsymbol r}_{\boldsymbol j}=(x_{\boldsymbol j},y_{\boldsymbol j},z_{\boldsymbol j})$. 
We start by studying the dynamics of a single excitation with wave function
{$\ket{\psi(t)}=\sum_{{\boldsymbol j}}c_{\boldsymbol j}(t)\sigma_{\boldsymbol j}^+\ket{\mathcal G}\ket{0}+\int d{\boldsymbol k}\sum_\lambda \psi_\lambda({\boldsymbol k},t)\ket{\mathcal G}\ket{{\boldsymbol k},\lambda}$}.
Here $\sigma_{\boldsymbol j}^+=\ket{e}_{\boldsymbol j}\!\bra{g}$, $\ket{\mathcal G}=\otimes_{\boldsymbol j}\ket{g}_{\boldsymbol j}$, $\ket{0}$ is the photonic vacuum state and $\ket{{\boldsymbol k},\lambda}$ the state with a single photon with wave vector ${\boldsymbol k}$ and polarization $\lambda$. We extend our results below to states with multiple excitations.

The atomic dynamics, due to successive photon emissions and reabsorptions, is obtained by integrating out the dynamics of the radiation modes $\psi_\lambda({\boldsymbol k},t)$ in a Born-Markov approximation. Assuming the field initially in the vacuum state $\psi_\lambda({\boldsymbol k},0)=0$, this yields
$\dot c_{\boldsymbol j}= -i  \sum_{{\boldsymbol j}'}\mathcal H_{{\boldsymbol j},{\boldsymbol j}'} c_{{\boldsymbol j}'}$, where in a frame rotating with $\omega_0$ \cite{Lehmberg1970,James1993,hecht},
\begin{equation}\label{eq:mathbbH}
\mathcal H_{{\boldsymbol j},{\boldsymbol j}'}\equiv-i(\gamma_e/2){\boldsymbol p}^{*}\cdot \hat{\boldsymbol G}({\boldsymbol r}_{{\boldsymbol j}}-{\boldsymbol r}_{\boldsymbol j'})\cdot{\boldsymbol p}
\end{equation}
is a non-hermitian effective Hamiltonian, whose hermitian part describes coherent exchanges of atomic excitations, while the non-hermitian part corresponds to dissipation accounting for radiation of photons.
 Here $\gamma_e$ is the spontaneous decay rate of each atom, and the dyadic Green's tensor ${\hat{\boldsymbol G}}({{\boldsymbol r}})$, representing the electric field
at position ${\boldsymbol r}$ generated by a dipole located at the origin, is the solution of $\boldsymbol \nabla\times \boldsymbol\nabla\times {\hat{\boldsymbol G}}({{\boldsymbol r}})-k_0^2 {\hat{\boldsymbol G}}({{\boldsymbol r}})+(6\pi i/k_0)\delta(\boldsymbol r)=0$ with {$\hat{\boldsymbol G}\big({\boldsymbol 0}\big)\equiv {\boldsymbol{\mathbb 1}}$}
accounting for independent single-atom decay (see details in \cite{SM}). The atomic transition polarization ${\boldsymbol p}$ is taken circular, with $z$ as quantization axis. 

\textit{Dark and bright eigenstates. ---}
The dynamics of atomic excitations, including their radiative properties, can be understood by studying the spectrum of $\mathcal H$. Denoting its eigenvalues as $\epsilon_{n}=\Delta_{n}-i\gamma_{n}/2$ (with $n=1,\ldots,2N$), $\Delta_{n}$ is interpreted as the self-energy of the collective atomic excitation given by the corresponding eigenstate $c_n$, while $\gamma_{n}$ is its spontaneous emission rate.
In particular, an eigenstate is \emph{subradiant} (or `dark') if spontaneous emission occurs with a rate suppressed below the single-atom decay rate $\gamma_e$. In view of the mirror symmetry of the system, all eigenstates have a definite parity, i.e., they can be written as $(c_n)_{({\boldsymbol j}_\perp,1)}=p_n (c_n)_{({\boldsymbol j}_\perp,2)}\equiv (v_n)_{{\boldsymbol j}_\perp}/\sqrt{2}$, with parity $p_n=\pm 1$. 

In Fig.~\ref{fig:setup}(d) we plot the decay rates [for the setup of Fig.~\ref{fig:setup}(c)], with  parameters chosen as explained below. One of the eigenstates is remarkably subradiant, with a decay rate of $\gamma_d\sim 10^{-3}\gamma_e$. We also represent the mean absolute value of the transverse quasi-momentum $\overline q$, which is obtained from the discrete Fourier transform of the corresponding eigenvectors $(\tilde v_n)_{{\boldsymbol q}}=\sum_{{\boldsymbol j}_\perp}(v_n)_{{\boldsymbol j}_\perp}e^{i\delta_\perp {\boldsymbol j}_\perp\cdot {\boldsymbol q}}/\sqrt{N}$ as $\overline{q}=\sum_{\boldsymbol q}|(\tilde v_n)_{{\boldsymbol q}}|^2|{\boldsymbol q}|$, with discrete quasi-momentum $\boldsymbol q=(q_x,q_y)$ where $q_{x,y}=-\pi/\delta_\perp+2\pi n_{x,y}/L_\perp$ ($n_{x,y}=0,1,...,N_\perp-1$). Two states have a distinctly low quasi-momentum $\overline q\ll k_0$: the dark state, as well as a `bright' state, which radiates photons with a rate $\gamma_b$ comparable to $\gamma_e$. We contrast our dark states with the $\overline q> k_0$ subradiant states in single layer setups,  studied e.g.~in Refs.~\cite{Asenjo-Garcia2017,Sutherland2016}.
In Fig.~\ref{fig:setup}(e) we show the probability amplitude of the eigenvectors $|(v_n)_{{\boldsymbol j}_\perp}|$ for the two states with lowest decay rates~\cite{SM}. 
 
\begin{figure}
\includegraphics[width=\columnwidth]{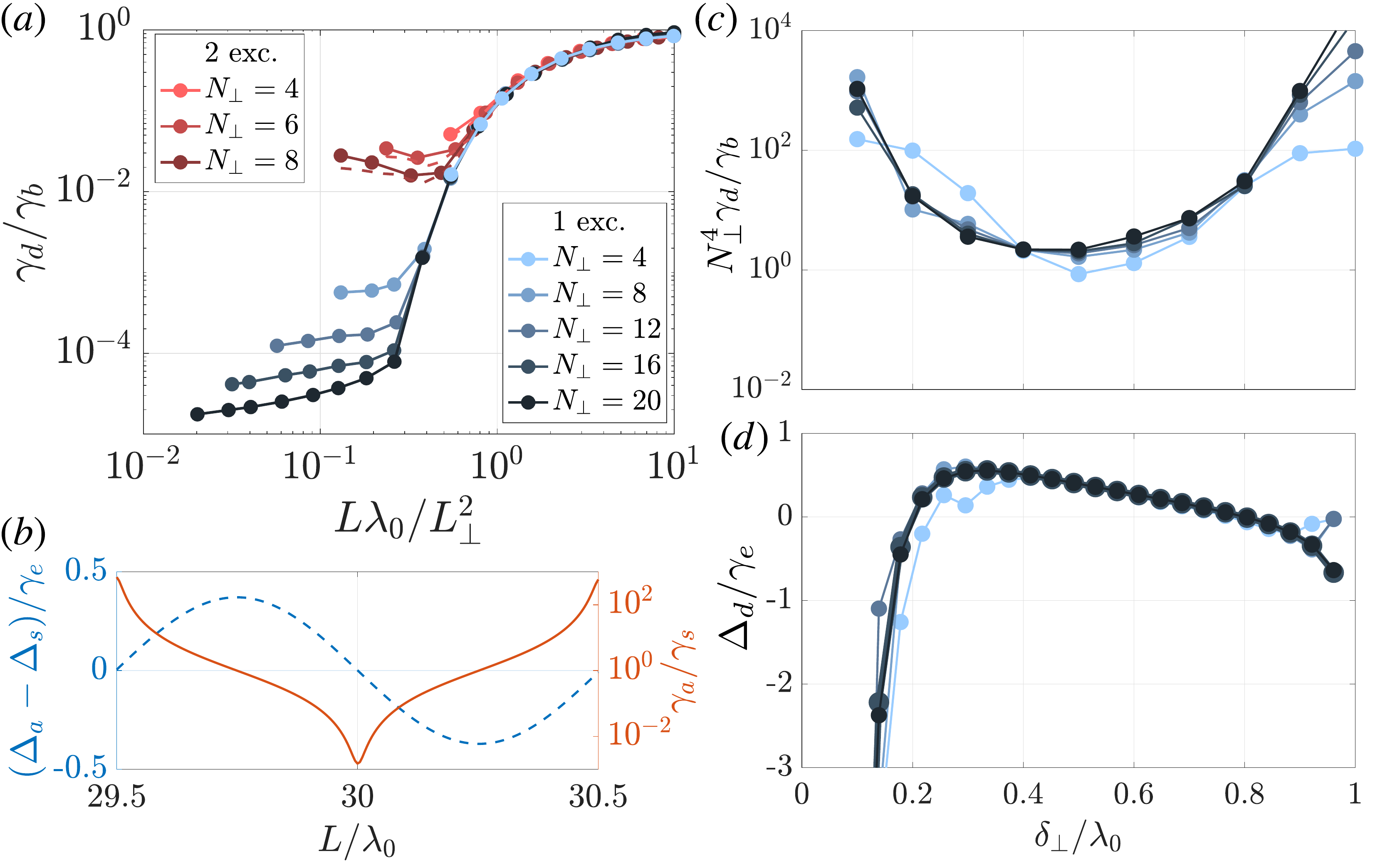} 
\caption{\textit{Dark and bright state properties.} (a) Ratio of dark and bright states decay rates for 1 (blue) and 2 (red) excitations, with {$\delta_\perp=\lambda_0/2$} and $L=m\lambda_0/2$ with integer $m$. (b)~Collective frequency shifts (dashed blue) and decay rates (red) of parity-symmetric ($s$) and anti-symmetric ($a$) single-excitation states, with $\delta_\perp=0.8\lambda_0$ and $N_\perp=12$. (c) Dark and bright state decay rates, and (d)~collective frequency shift of the dark state, for $L=2\lambda_0$, $\delta_\perp=\lambda_0/2$, and $N_\perp$ as in (a) for 1 excitation.
\label{fig:fig2}}
\end{figure}

This pair of dark and bright states can be understood by considering first the situation where the arrays are infinite ($N_\perp\to\infty$), and the eigenstates are plane waves $(v_n)_{{\boldsymbol j}_\perp}=e^{i\delta_\perp {\boldsymbol j}_\perp\cdot {\boldsymbol q}_n}/\sqrt{N}$ with continuous quasi-momentum ${\boldsymbol q}_n$. We now make two assumptions: First, the lattice spacing satisfies $\delta_\perp<\lambda_0$. Under this condition, we obtain, provided $|{\boldsymbol q}_n|\leq 2\pi/\delta_\perp - k_0$ \cite{SM},
\begin{equation}
\gamma_n=\Gamma[1+p_n\cos(k_z L)]\tfrac{k_z^2+|{\boldsymbol q}_n|^2/2}{k_0 q_z},\label{eq:gammanplanewaves}
\end{equation} 
with $\Gamma=3\pi\gamma_e/(k_0\delta_\perp)^2$, $k_z=\sqrt{k_0^2-|{\boldsymbol q}_n|^2}$. Considering in particular the {symmetric} ($p_n=1$) and {antisymmetric} ($p_n=-1$) eigenstates with ${\boldsymbol q}_n={\boldsymbol 0}$, we obtain a pair of states with decay rates $\gamma_{s/a}=\Gamma[1\pm \cos(k_0L)]$. Similarly, their self-energies are $\Delta_{s/a}=\pm(\Gamma/2)\sin(k_0L)+\Delta_d$, as depicted in Fig.~\ref{fig:fig2}(b), where $\Delta_d$ is a collective Lamb shift evaluated numerically. Our second assumption is that $k_0L=m\pi$ with integer $m$, so that either $\gamma_s$ or $\gamma_a$ vanishes due to interference in the emission of the two arrays, while the other reduces to $\gamma_b=2\Gamma$. 
The corresponding Bell states 
\begin{equation}
\ket{\psi_{d/b}}=\frac{1}{\sqrt{2N}}\sum_{\boldsymbol j_\perp}\left[\sigma^+_{(\boldsymbol j_\perp,1)}\mp(-1)^m\sigma^+_{(\boldsymbol j_\perp,2)}\right]\ket{\mathcal G}
\end{equation}
are thus respectively `dark' and `bright'. 

For finite-sized arrays, the eigenstates $(v_n)_{{\boldsymbol j}_\perp}$ are confined, which has two consequences yielding a finite decay rate $\gamma_d$ for the dark state. First, photon emission in transverse directions is not perfectly cancelled. Second, photons emitted along $z$ have a finite spread of transverse momentum, and thus diffract when propagating between the two arrays, thereby hindering the interference of emission. This can be mitigated by curving the arrays according to the phase profile of a Gaussian mode $\mathcal E({\boldsymbol r})$ propagating along $z$ [as shown in Fig.~\ref{fig:setup}(c)], in analogy to the mirrors of an optical cavity. As represented in Fig.~\ref{fig:setup}(e), the spatial distribution of the dark state (as well as the bright state) is then $(v_d)_{{\boldsymbol j}_\perp}\propto \mathcal E({\boldsymbol r}_{({\boldsymbol j}_\perp,1)})$ \cite{paraxial}. Alternatively, one can add optical elements between the arrays, such as lenses or fibers.

The spatial profile of the electric field, generated by (virtual) photon exchanges between the atomic dipoles in the dark state, reads ${\boldsymbol \psi}({\boldsymbol r})\sim \sum_{\boldsymbol j}c_{\boldsymbol j} \hat {\boldsymbol G}({\boldsymbol r}-{\boldsymbol r}_{\boldsymbol j})\cdot {\boldsymbol p}$, and forms a standing wave [see Fig.~\ref{fig:setup}(c)]. We emphasize that -- although the system resembles a cavity with each array acting as a mirror -- we are interested here in the quantum state of the \emph{atoms}. More precisely, the ratio of atomic to photonic excitations in the dark state is given by $\Gamma L /(2c)$ with speed of light $c$ \cite{SM}, which is assumed negligible when integrating the field dynamics above, amounting to neglecting retardation effects in the atomic dynamics. This is in analogy to atomic cavities built from strings of atoms coupled to a 1D waveguide \cite{ChangDEandJiangLandGorshkovAVandKimble2012,PhysRevA.93.023808}.

We now discuss how the geometric parameters ($N_\perp, L, \delta_\perp$) affect the spectral properties of the system.
In Fig.~\ref{fig:fig2}(a) we show the scaling of $\gamma_d/\gamma_b$ as the relevant figure of merit, with the waist of $\mathcal E({\boldsymbol r})$ minimizing this ratio. Low ratios can be achieved for  $L\lesssim L_\perp^2/\lambda_0$, a condition set by the diffraction limit, i.e.~the spot size of the Gaussian mode must be smaller than the surface of the arrays. Remarkably, this condition allows to achieve strong subradiance even when the characteristic size of each array $L_\perp$ is much smaller than their separation $L$, i.e., the subradiant state is `non-local'.
As an example, for $N_\perp=20$ and $\delta_\perp= 0.8\lambda_0$ (i.e., $L_\perp=16\lambda_0$), we obtain $\gamma_d/\gamma_b\sim10^{-2}$ for $L\sim 130\lambda_0$. 

In Fig.~\ref{fig:fig2}(b) we observe that the interference mechanism is quite sensitive to the separation between arrays, as small deviations of $L$ compared to $\lambda_0$ will greatly increase the decay rate $\gamma_d$ [see Eq.~\eqref{eq:gammanplanewaves}]. 
In Fig.~\ref{fig:fig2}(c) we show the effect of the lattice spacing on the saturation value of Fig.~\ref{fig:fig2}(a) for small $L$. The ratio of dark to bright state decay rates is minimal for $\delta_\perp=\lambda_0/2$, for which the emission in transverse directions is best cancelled, and scales with the atom number as $\gamma_d/\gamma_b\sim 1/N_\perp^4$. 
The collective shift $\Delta_d$ on the other hand is typically of the order of $\gamma_e$ [c.f.~Fig.~\ref{fig:fig2}(d)]. It can be positive or negative depending on $\delta_\perp$, and vanishes around $\delta_\perp=0.2\lambda_0$ and $\delta_\perp=0.8\lambda_0$ (see also Ref.~\cite{Shahmoon2017}). 

\begin{figure}
\includegraphics[width=\columnwidth]{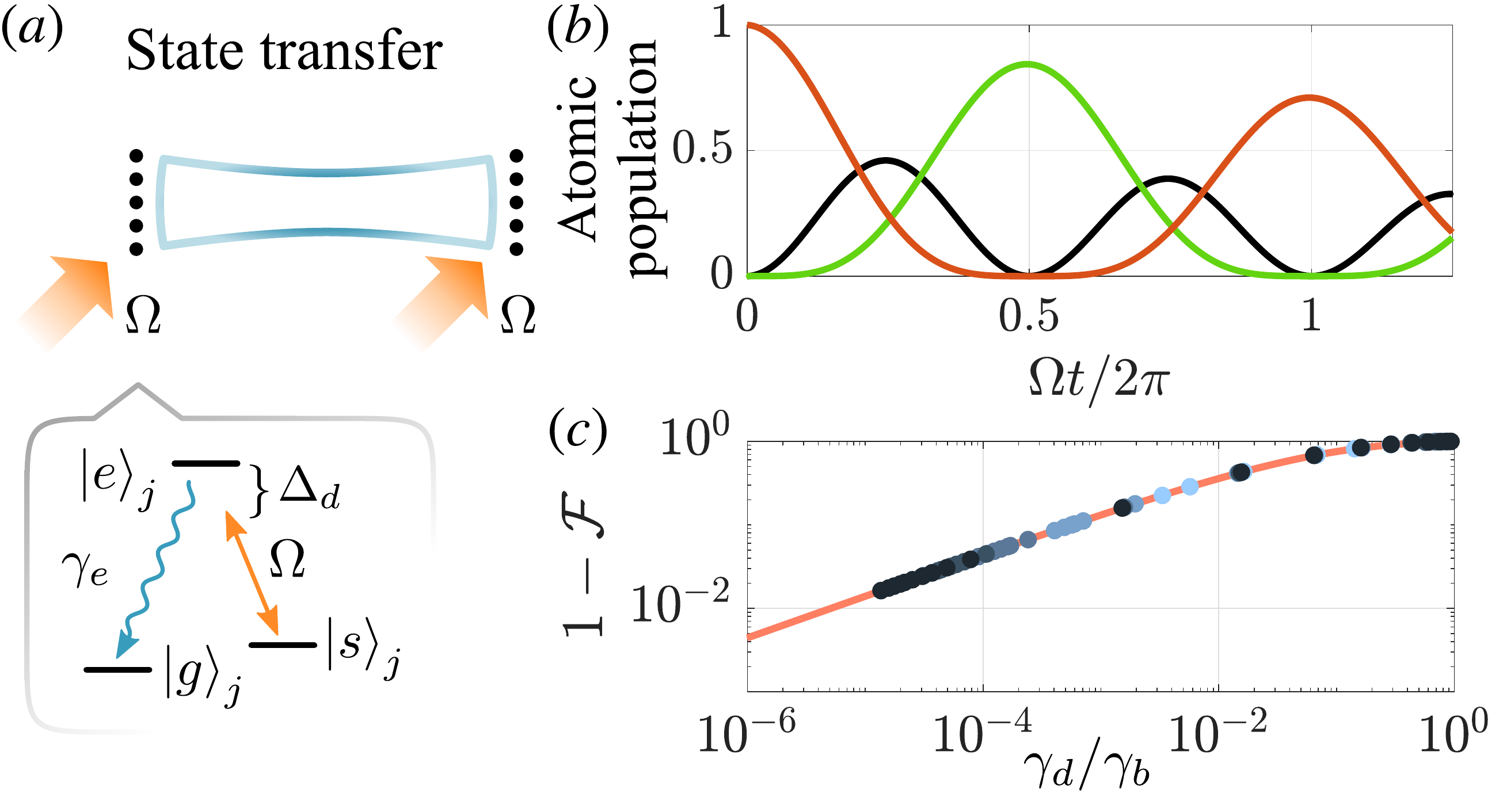} 
\caption{\textit{Quantum state transfer between `local' quantum memories.} (a)~Sketch and atomic $\Lambda$-level structure for coupling quantum memories. A weak homogeneous field $\Omega$, resonant with the collective atomic shift $\Delta_d$, drives the $\ket{e}\to\ket{s}$ transition. (b)~Temporal evolution of the atomic populations for the initial state $S_1^+\ket{\mathcal G}$, with $N_\perp=12$, $L=30\lambda_0$, $\delta_\perp=0.8\lambda_0$. Red (green): number of atoms in state $\ket{s}$ in the first (second) array. Black: total number of atoms in state $\ket{e}$. (c)~Infidelity for quantum state transfer as function of dark and bright state decay rates. Blue dots: parameters of Fig.~\ref{fig:fig2}(a) for 1 excitation. Red curve: Eq.~\eqref{eq:Fopt}. \label{fig:fig3}}
\end{figure}

\textit{Dark state preparation and quantum state transfer.~---}
In order to prepare the atoms in the dark state, we consider the setup represented in Fig.~\ref{fig:fig3}(a), where the atomic level structure now includes a third state $\ket{s}$. We assume that the system is initially in a superposition state of the first array {$S_1^+\ket{\mathcal G}$}, with $S_1^+=\sum_{{\boldsymbol j}_\perp}(v_d)_{{\boldsymbol j}_\perp} \ket{s}_{({\boldsymbol j}_\perp,1)}\!\bra{g}$. 
This could be realized for instance using laser-dressed Rydberg-Rydberg interactions \cite{Grankin2018,PhysRevLett.121.123605}, or single photon pulses \cite{SM}.
Moreover, we assume a coherent field drives the $\ket{s}\to\ket{e}$ transition in the first array with Rabi frequency $\Omega$, resonantly with the collective shift $\Delta_d$. The atoms are thus driven from state $S_1^+\ket{\mathcal G}$ to a superposition of dark and bright states $\sum_{{\boldsymbol j}_\perp}(v_d)_{{\boldsymbol j}_\perp}\sigma_{({\boldsymbol j}_\perp,1)}^+\ket{\mathcal G}=(1/\sqrt{2})(\sigma_b^++\sigma_d^+)\ket{\mathcal G}$, where the operators $\sigma^+_{d/b}\equiv\sum_{{\boldsymbol j}_\perp}(v_d)_{{\boldsymbol j}_\perp}(\sigma^+_{({\boldsymbol j}_\perp,1)}\mp [-1]^m \sigma^+_{({\boldsymbol j}_\perp,2)})/\sqrt{2}$ create a dark/bright atomic excitation.
The decay rate of bright excitations can be orders of magnitude larger than for dark excitations, such that their contribution to the dynamics is vastly different. If $\gamma_b\gg\Omega$, the bright mode can be adiabatically eliminated, and contributes an effective loss with rate $\Omega^2/\gamma_b$, which can vanish in the spirit of a quantum Zeno effect. On the other hand, if $\Omega\gg \gamma_d$, the dynamics will yield oscillations between the initial state and the {non-local} dark state.

This mechanism can be exploited for quantum state transfer between the two arrays. Here, an initial qubit superposition state in the first array $\ket{\psi_i}=c_g\ket{\mathcal G}+c_s S_1^+\ket{\mathcal G}$ (with $|c_g|^2+|c_s|^2=1$) is transferred deterministically to the second array. That is, we realize the process {$\ket{\psi_i}\to\ket{\psi_f}=c_g\ket{\mathcal G}+c_s S_2^+\ket{\mathcal G}$}, where {$S_2^+=\sum_{{\boldsymbol j}_\perp} (v_d)_{{\boldsymbol j}_\perp} \ket{s}_{({\boldsymbol j}_\perp,2)}\!\bra{g}$}, with high fidelity $\mathcal{F}\approx1$~\cite{footnote1}. By driving atoms in both arrays with Rabi frequency $\Omega$, the state $S_2^+\ket{\mathcal G}$ is coupled to the opposite superposition {$\sum_{{\boldsymbol j}_\perp}(v_d)_{{\boldsymbol j}_\perp}\sigma_{({\boldsymbol j}_\perp,2)}^+\ket{\mathcal G}=(1/\sqrt{2})(\sigma_b^+-\sigma_d^+)\ket{\mathcal G}$}, and we can write an effective model, where the system is described by four excitation modes: two `local' modes, with creation operators $S_1^+$ and $S_2^+$, which represent quantum memories in $\ket{\psi_i}$ and $\ket{\psi_f}$; and two `non-local' bright and dark modes, with creation operators $\sigma_b^+$ and $\sigma_d^+$, connecting the two memories. The dynamics can then be described by a Lindblad master equation for the density matrix of the atoms $\rho$, as {$\dot\rho=-i[H_\text{eff},\rho]+\gamma_d \mathcal D[\sigma_d^-]\rho + \gamma_b \mathcal D[\sigma_b^-]\rho$}, where {$\mathcal D[a]\rho\equiv a\rho a^\dagger -(1/2)(a^\dagger a\rho+\rho a^\dagger a)$}, and with an effective Hamiltonian
\begin{equation}\label{eq:Heff}
H_\text{eff}=\frac{\Omega}{\sqrt{2}}\left[\sigma_b^+(S_1^-+S_2^-)+ \sigma_d^+(S_1^--S_2^-) \right]+\text{h.c.}
\end{equation}

The evolution of the system is shown in Fig.~\ref{fig:fig3}(b), demonstrating transfer at time $t=\pi/\Omega$~\cite{fn}. We emphasize that our protocol does not require tailoring the temporal shape of exchanged photons, in contrast to deterministic quantum state transfer protocols with `flying' photonic qubits \cite{Cirac1997, Grankin2018}. Fig.~\ref{fig:fig3}(c) represents in red the optimal achievable fidelity for given $\gamma_{d,b}$, which reads
 \begin{equation}\label{eq:Fopt}
 \mathcal{F}\approx e^{-\pi\sqrt{2\gamma_d/\gamma_b}},
 \end{equation} 
showing the requirement $\gamma_b\gg\gamma_d$. The blue dots represent simulations for atomic arrays with the parameters of Fig.~\ref{fig:fig2}(a), with the optimal drive given by $\Omega= \sqrt{\gamma_d\gamma_b/8}$~\cite{SM}. 
 As noted above, our treatment neglects effects of retardation in atomic dynamics;  Eq.~\eqref{eq:Fopt} remains, however, valid even for large delay times, although at the cost of a slowdown of the dynamics~\cite{SM}.

\textit{Probing the dark state. ---} 
The existence of the dark state can be detected in the reflection of an external laser (see details in \cite{SM}). We consider here a weak probing field with frequency $\omega_0+\Delta_d$, propagating along $z$ in the Gaussian mode $\mathcal E({\boldsymbol r})$, and driving atoms prepared in the ground state $\ket{\mathcal G}$. Assuming the transition frequency of the atoms in each array is additionally detuned, by $\Delta$ for atoms in the first array and either $\Delta$ or $-\Delta$ for the second array, the dark and bright states are then revealed in the width of the resonance peak of the reflectivity $R(\Delta)$. We obtain $R= (\gamma_b-\gamma_d)^2/(\gamma_b^2+4\Delta^2)$ for symmetric detuning, and $R = (\gamma_b-\gamma_d)^2/(\gamma_b+4\Delta^2/\gamma_d)^{2}$ for opposite detuning, which both have a peak at $\Delta=0$~\cite{SM}; the widths of these peaks are given by $\gamma_b$ and $\sim\!\sqrt{\gamma_d\gamma_b}$, respectively, allowing for a direct probing of the dark state lifetime.

\textit{Experimental considerations. ---}
The level structure can be implemented in neutral atoms using for instance stretched states of ${}^{87}$Rb for {$\ket{g}=\ket{5S_{1/2},F=2,m_F=2}$} and {$\ket{e}=\ket{5P_{3/2},F=3,m_F=3}$}, along with a strong magnetic field to eliminate other hyperfine states from the dynamics. The level $\ket{s}$ needs to be coherently coupled to the excited state, while avoiding spontaneous decay from $\ket{e}$ to $\ket{s}$. This could be realized for example using a Rydberg state $\ket{s}=\ket{nS_{1/2},m=1/2}$, with higher energy \cite{Manzoni2017}, or another ground state $\ket{s}=\ket{5S_{1/2},F=1,m_F=1}$, coupled to $\ket{e}$ via a two-photon transition \cite{PhysRevA.78.053816}.
Alternatively, one can use for the optical transition atoms with a $J=0\to J=1$ transition, e.g. ${}^{88}$Sr; while this introduces three excited states with orthogonal dipole matrix elements, our results for dark and bright state decay rates remain qualitatively similar \cite{SM}.

The atomic trap is characterized by a finite temperature and Lamb-Dicke parameter $\eta$ \cite{Ludlow2015}. The resulting spread of the atomic wavefunction yields a renormalization of the decay rates as
{$\gamma_{d/b}\to  \gamma_{d/b}[1- \eta^2 (2n_\text{th}+1)]+\gamma_e\eta^2(2n_\text{th}+1)$}~\cite{SM},
where $n_\text{th}$ is the thermal occupation number of trap states, and we assumed $\eta\sqrt{2n_\text{th}+1}\ll 1$ and $\gamma_e \eta \sqrt{2n_\text{th}+1}\ll \omega_\nu$, with $\omega_\nu$ the atomic motional frequency. We thus need {$\eta^2(2n_\text{th}+1)\lesssim \gamma_d/\gamma_e$}.
The effect of missing atoms is similar~\cite{SM}; for a defect probability $p$, we find 
{$\gamma_{d/b}\to  \gamma_{d/b}(1- p)+\gamma_e p+\mathcal O(p^2)$}, i.e.~we require $p\lesssim\gamma_d/\gamma_e$.


\textit{Multiple excitations. ---}
For states with multiple excitations, the dynamics can be studied again by analyzing the spectral properties of the non-hermitian effective Hamiltonian, which now takes the form $H_\text{dip}=\sum_{{\boldsymbol j},{\boldsymbol j}'} \mathcal H_{{\boldsymbol j},{\boldsymbol j}'} \sigma_{\boldsymbol j}^+\sigma_{{\boldsymbol j}'}^-$ \cite{SM}.  Since each atom cannot be excited more than once,  the doubly-excited state $(\sigma_d^+)^2\ket{\mathcal G}$ cannot be an exact eigenstate of $H_\text{dip}$. An analytical expression for the resulting decay rates can, however, be obtained by treating the non-linearity as perturbation, where each excitation effectively acts as a defect for the other, with the `defect' probability $p$ identified as the inverse participation ratio $p=\sum_{{\boldsymbol j}_\perp}|(v_d)_{{\boldsymbol j}_\perp}|^4$ (see Ref.~\cite{SM}). In Fig.~\ref{fig:fig2}(a) we show in red, for the eigenstate closest to $(\sigma_d^+)^2\ket{\mathcal G}$, the ratio of the decay rate per excitation $\gamma_d^{(2)}$ and $\gamma_b$, which is well captured by this analytical approximation (dashed red curves). 

For large $N_\perp$, we thus expect {$\gamma_d^{(2)}\sim\gamma_e/N_\perp^2$}, since $(v_d)_{{\boldsymbol j}_\perp}\sim 1/N_\perp$.  
Two regimes can then be explored. First, for {$\gamma_d,\gamma_d^{(2)} \ll \gamma_b$} the system becomes effectively almost linear, and in particular the protocol for quantum state transfer above remains valid, with the replacement $\gamma_d\to\gamma_d^{(2)}$. This can be used to transfer states with more than one excitation, e.g. quantum error correcting states such as cat or binomial states \cite{Michael2016}, allowing in principle to reach fidelities beyond Eq.~\eqref{eq:Fopt}. Second, if {$\gamma_d\ll\gamma_d^{(2)}, \gamma_b$}, excitations of radiating two-excitation states can be adiabatically eliminated, exploiting again the quantum Zeno effect. 
This mechanism can be used to effectively block the transfer from the memories to the dark state, and thereby can operate as a controlled-phase gate \cite{Dzsotjan2010}. Moreover, by the same principle, weakly driving the optical transition of atoms in one of the arrays generates Rabi oscillations between $\ket{\mathcal G}$ and $\sigma_d^+\ket{\mathcal G}$ as a two-level system, which can also be used to prepare the system in the dark state, e.g. for entanglement generation between memories, or as single-photon source.

\textit{Conclusion. ---} We have shown that distant single-layered arrays of two-level atoms can support subradiant (long-lived) states as collective excitations in the form of Bell superpositions.
Our setup constitutes a building block for a modular quantum architecture, where quantum information, stored and processed in atomic arrays, is exchanged via dark modes. Moreover, the separation between arrays can be drastically increased by adding lenses or optical fibers to mediate photons between the arrays, although at the cost of adding decoherence channels.
While we discussed here implementations with atoms in optical lattices, our results remain valid for other types of emitters, including for instance in solid-state platforms such as color centers in diamond \cite{Doherty2013}, quantum dots \cite{Lodahl2015}, or
monolayers of transition metal dichalcogenides \cite{Zhou}.

\begin{acknowledgments}
We thank A.~Asenjo-Garcia, D.~Chang, F.~Robicheaux, J.~Ruostekoski and M.~Saffman for comments on the manuscript. This work was supported by the Army Research Laboratory Center for Distributed Quantum Information via the project SciNet, the ERC Synergy Grant UQUAM and the SFB FoQuS (FWF Project No. F4016-N23). 
\end{acknowledgments}

\bibliography{subradiance}

\begin{thebibliography}{42}
\bibitem [{par()}]{paraxial}%
  \BibitemOpen
\bibfield  {journal} {  }\href@noop {} {\bibinfo  {journal} {Analytical
  expressions are obtained below within a paraxial approximation}\
  }\BibitemShut {NoStop}%
\bibitem [{foo()}]{footnote1}%
  \BibitemOpen
  \href@noop {} {\bibinfo  {journal} {We define the fidelity
  $\mathcal{F}\equiv\text{Tr}\left({\rho}(T)\ket{\psi_f}\!\bra{\psi_f}\right)$,
  as the probability of successful transfer, with $\rho(T)$ the density matrix
  at the end of the protocol with $c_s=1$}\ }\BibitemShut {NoStop}%
\bibitem [{fn()}]{fn}%
  \BibitemOpen
\bibfield  {journal} {  }\href@noop {} {\bibinfo  {journal} {The protocol is
  terminated by switching off the laser drive}\ }\BibitemShut {NoStop}%
\end{thebibliography}

\newpage
\newpage 

\onecolumngrid
\newpage
{
\center \bf \large 
Supplemental Material for: \\
Subradiant Bell states in distant atomic arrays\vspace*{0.1cm}\\ 
\vspace*{0.0cm}
}
\begin{center}
P.-O.~Guimond, A.~Grankin, D.\,V.~Vasilyev, B.~Vermersch and P.~Zoller\\
\vspace*{0.15cm}
\small{\textit{Center for Quantum Physics,
University of Innsbruck, Innsbruck A-6020, Austria\\
and Institute for Quantum Optics and Quantum Information, Austrian Academy of Sciences, Innsbruck A-6020,
      Austria}}\\
\vspace*{0.25cm}
\end{center}

\twocolumngrid

\section{Quantum optical model}\label{sec:opticalmodel}
Here we provide details on our model and the definitions in Eq.~\eqref{eq:mathbbH}. 
We first consider the full system comprising the atoms in the first array (labeled with $j_z=1$), in the second array ($j_z=2$), and including the electromagnetic field. Each atom has a ground state $\ket{g}$ and an excited state $\ket{e}$. The dynamics is governed by the Hamiltonian
$H_\text{tot}=H_a+H_f+V_{af},$
where $H_a$ acts on the atoms, and reads ($\hbar=1$)
\begin{equation*}
H_a=\sum_{\boldsymbol j}\omega_0 \sigma_{\boldsymbol j}^+\sigma_{\boldsymbol j}^-+H_\text{drive},
\end{equation*} 
with $\boldsymbol j=(\boldsymbol j_\perp,j_z)$, $\boldsymbol j_\perp=(j_x,j_y)$ and $1\leq j_x,j_y \leq N_\perp$.
Here $\omega_0$ is the atomic transition frequency, $\sigma_{\boldsymbol j}^+\equiv\ket{e}_{\boldsymbol j}\bra{g}$, and $H_\text{drive}$ is an additional term accounting for possible additional laser drivings. The electromagnetic field Hamiltonian reads $H_f=\int d{\boldsymbol k} \sum_\lambda \omega_{{\boldsymbol k}} b^\dagger_{\lambda,{\boldsymbol k}}b_{\lambda, {\boldsymbol k}}$, where $\omega_{{\boldsymbol k}}=c | {\boldsymbol k}|$ with $c$ the speed of light, $b_{\lambda,{\boldsymbol k}}$ is the annihilation operator for photons with helicity $\lambda=\pm 1$ satisfying $[b_{\lambda,{\boldsymbol k}},b^\dagger_{\lambda',{\boldsymbol k}'}]=\delta_{\lambda,\lambda'}\delta({\boldsymbol k}-{\boldsymbol k}')$. Finally, the interaction Hamiltonian reads
\begin{equation*}
V_{af}=-d\sum_{\boldsymbol j}(\sigma_{\boldsymbol j}^+{\boldsymbol p}^*+\sigma_{\boldsymbol j}^-{\boldsymbol p})\cdot \hat {\boldsymbol E}({\boldsymbol r}_{\boldsymbol j})+\text{h.c.},
\end{equation*}  
where $d$ is the atomic dipole, ${\boldsymbol p}$ the atomic transition polarization (we assume circular polarization), and the electric field operator expresses as 
\begin{equation*}
\hat {\boldsymbol E}({\boldsymbol r})=i\int d{\boldsymbol k}\sum_{\lambda} \epsilon_{|{\boldsymbol k}|} b_{\lambda,{\boldsymbol k}} e^{i {\boldsymbol k}\cdot {\boldsymbol r}} {\boldsymbol e}_{\lambda,{\boldsymbol k}},
\end{equation*}
with ${\boldsymbol e}_{\lambda,{\boldsymbol k}}$ the polarization unit vector, \mbox{$\epsilon_k=\sqrt{ck/(2[2\pi]^3\varepsilon_0)}$}, and $\varepsilon_0$ the vacuum permittivity. 

Assuming the electromagnetic field is initially in the vacuum state, the field dynamics can be integrated, to obtain a Lindblad master equation for the reduced system of the atoms, within a Born-Markov approximation. We obtain \cite{Lehmberg1970,hecht}
\begin{equation}
\begin{aligned}
\label{eq:ME2nodes}
\frac{d\rho}{dt}= -i\Big[H_\text{drive}  +\text{Re}(H_\text{dip}),\rho\Big]-2&\mathcal L(\rho).
\end{aligned}
\end{equation}
The resulting non-hermitian dipole-dipole interaction Hamiltonian reads $H_\text{dip}=\sum_{{\boldsymbol j},{\boldsymbol j}'}\mathcal H_{{\boldsymbol j},{\boldsymbol j}'}\sigma_{\boldsymbol j}^+\sigma_{{\boldsymbol j}'}^-$, where 
\begin{equation}\label{mathcalH}
\mathcal H_{{\boldsymbol j},{\boldsymbol j}'}=-i(\gamma_e/2){G}({{\boldsymbol r}_{\boldsymbol j}-{\boldsymbol r}_{\boldsymbol j'}}),
\end{equation} 
with $\gamma_e=k_0^3 d^2/(3\pi\varepsilon_0)$ the single-atom decay rate, $G({\boldsymbol r})={\boldsymbol p}^*\cdot\hat {\boldsymbol G}({\boldsymbol r})\cdot {\boldsymbol p}$, and the dyadic Green's tensor taking the explicit form 
\begin{equation*}
\begin{aligned}{\hat{\boldsymbol G}}({\boldsymbol r})=\frac{3e^{ik_{0}r}}{2i(k_{0}r)^{3}}\Big[ & \left((k_{0}r)^{2}+ik_{0}r-1\right)\\
 & +\left(-(k_{0}r)^{2}-3ik_{0}r+3\right)\frac{{\boldsymbol r}\otimes{\boldsymbol r}}{r^{2}}\Big],
\end{aligned}
\end{equation*}
which represents the field at position ${\boldsymbol r}$ emitted by a dipole at the origin, with
\begin{equation}\label{eq:maxwell}
\boldsymbol{\nabla}\times\boldsymbol \nabla\times \hat{\boldsymbol G}(\boldsymbol r)-k_0^2 \hat{\boldsymbol G}(\boldsymbol r)=-\frac{6\pi i}{k_0}\delta(\boldsymbol r), 
\end{equation}
and we define $\hat{\boldsymbol G}({\boldsymbol 0})={\boldsymbol{ \mathbb 1}}$. 
The last term in Eq.~\eqref{eq:ME2nodes} reads
\begin{equation*}
\mathcal L(\rho)=\sum_{{\boldsymbol j},{\boldsymbol j}'}\text{Im}(\mathcal H)_{{\boldsymbol j},{\boldsymbol j}'} \left( \sigma_{{\boldsymbol j}'}^-\rho \sigma_{{\boldsymbol j}}^+-\frac12\left\{\sigma_{{\boldsymbol j}}^+\sigma_{{\boldsymbol j}'}^-,\rho\right\}\right).
\end{equation*}
In writing the master equation we moved to a rotating frame with the atomic transition frequency $\omega_0$, and we made the following assumptions. (i)~Rotating wave approximation: counter-rotating terms (such as $\sigma_{\boldsymbol j}^+\sigma_{{\boldsymbol j}'}^+$), which do not preserve the number of atomic excitations, are neglected. (ii)~Markov approximation: retardation effects due to finite light velocity are also neglected. 

We can notice from Eq.~\eqref{eq:ME2nodes} that the collective emission properties of the arrays are determined by the spectrum of $H_\text{dip}$. Since $H_\text{dip}$ conserves the total number of atomic excitations $\sum_{\boldsymbol j}\sigma_{\boldsymbol j}^+\sigma_{\boldsymbol j}^-$, we can evaluate its eigenstates in each excitation subspace separately. In particular, for single excitation eigenstates this amounts to diagonalizing $\mathcal H$. Let us assume the atomic arrays are initially prepared in one of these eigenstates $\ket{\psi_n}$, with complex eigenvalue $\epsilon_n=\Delta_n-i\gamma_n/2$, containing $N_\text{exc}$ atomic excitations, i.e. $\rho(0)=\ket{\psi_n}\bra{\psi_n}$. Assuming here $H_\text{drive}=0$, the dynamics of Eq.~\eqref{eq:ME2nodes} will yield 
\begin{equation*}
\rho(t)= e^{-\gamma_nt}\ket{\psi_n}\bra{\psi_n}+\rho'(t),
\end{equation*}
where $\rho'(t)$ contains strictly less than $N_\text{exc}$ atomic excitations, and $\gamma_n$ is thus interpreted as the decay rate of the eigenstate $\ket{\psi_n}$. More generally, starting from an initial mixture on the subspace with $N_\text{exc}$ excitations, we can write 
\begin{equation*}
\rho(t)=\rho_{N_\text{exc}}(t)+\rho'(t),
\end{equation*}
 where $\rho_{N_\text{exc}}(t)$ is a density matrix with $N_\text{exc}$ excitations satisfying
 \begin{equation*}
 \frac{d\rho_{N_\text{exc}}}{dt}=-i H_\text{dip} \rho_{N_\text{exc}}+i\rho_{N_\text{exc}}H_\text{dip}^\dagger.
 \end{equation*}
 
 \section{Spectral analysis of $\mathcal H$}
 Here we discuss the spectrum of $\mathcal H$ (i.e., the spectrum of $H_\text{dip}$ in the single-excitation subspace), in the cases of infinite and finite arrays. We then explain how the system can be probed with a laser to measure the bright and dark state decay rates. 
 
\subsection{Infinite planar arrays}\label{sec:infinitearrays}
We first derive analytical expressions for the spectrum of $\mathcal H$ in the case of infinite planar arrays ($N_\perp\to\infty$). We use the identity \cite{hecht}
\begin{equation*}
\frac{e^{ik_0r}}{r}=\frac{i}{2\pi}\int d{\boldsymbol q}\frac{e^{i{\boldsymbol q}\cdot {\boldsymbol r}_\perp}e^{i q_z |z|}}{q_z},
\end{equation*}
where ${\boldsymbol r}=({\boldsymbol r}_\perp,z)$, and $q_z=\sqrt{k_0^2-|{\boldsymbol q}|^2}$. This allows us to rewrite the Green's tensor as
\begin{equation}\label{eq:Gintdq}
\hat {\boldsymbol G}({\boldsymbol r})=\frac{3}{4\pi k_0^3}\int d{\boldsymbol q}\left[k_0^2 {\boldsymbol {\mathbb 1}}-{\boldsymbol {\overline Q}}\otimes {\boldsymbol {\overline Q}} \right]\frac{e^{i{\boldsymbol q}\cdot {\boldsymbol r}_\perp}e^{i q_z |z|}}{q_z},
\end{equation}
where ${\boldsymbol {\overline Q}}=({\boldsymbol q},q_z \text{sgn}(z))$.

Due to translational invariance and parity symmetry, the eigenstates of $\mathcal H$ are plane waves 
\begin{equation*}
(c_n)_{\boldsymbol j}=e^{i \delta_\perp {\boldsymbol j}_\perp\cdot{\boldsymbol q}_n}e^{i\pi \frac{p_n-1}{2}(j_z-1)}/\sqrt{2N},
\end{equation*} 
with ${\boldsymbol q}_n$ in the first Brillouin zone and $p_n=\pm1$ the eigenstate parity. Next we use the relation
\begin{equation}\label{eq:sumgdef}
\sum_{{\boldsymbol j}_\perp} e^{i \delta_\perp  {\boldsymbol j}_\perp\cdot{\boldsymbol q}}=\left(\frac{2\pi}{\delta_\perp}\right)^2 \sum_{{\boldsymbol g}}\delta^{(2)}({\boldsymbol q}-{\boldsymbol g}),
\end{equation}
where the sum on the right-hand side runs over vectors ${\boldsymbol g}$ of the reciprocal lattice, i.e., $g_{x,y}=(2\pi/\delta_\perp)m_{x,y}$ with integer $m_{x,y}$. We thus obtain from Eq.~\eqref{mathcalH}
\begin{widetext}
\begin{equation}\label{eq:eigHexplicit}
\begin{aligned}
&\sum_{{\boldsymbol j}'}\mathcal H_{{\boldsymbol j},{\boldsymbol j}'}(c_n)_{{\boldsymbol j}'}\\
&=-\frac{3i\gamma_e}{8\pi k_0^3}\sum_{{\boldsymbol j}'_\perp,j'_z}\int d{\boldsymbol q}e^{i\delta_\perp {\boldsymbol j}_\perp\cdot{\boldsymbol q}}\left[k_0^2-|{\boldsymbol q}\cdot {\boldsymbol p}|^2 \right]\frac{e^{i\delta_\perp {\boldsymbol j}'_\perp\cdot({\boldsymbol q}_n-{\boldsymbol q})}e^{i q_z |z_{\boldsymbol j}-z_{\boldsymbol j'}|}}{q_z\sqrt{2N}}e^{i\pi \frac{p_n-1}{2}(j'_z-1)}\\
&=-(c_n)_{\boldsymbol j}\frac{3(2\pi)^2i\gamma_e}{8\pi k_0(k_0\delta_\perp)^2}\sum_{j'_z}\sum_{\boldsymbol g}\left[k_0^2-|({\boldsymbol q}_n-{\boldsymbol g})\cdot {\boldsymbol p}|^2 \right]\frac{e^{i \sqrt{k_0^2-|{\boldsymbol q}_n-{\boldsymbol g}|^2} |z_{\boldsymbol j}-z_{\boldsymbol j'}|}}{\sqrt{k_0^2-|{\boldsymbol q}_n-{\boldsymbol g}|^2}}e^{i\pi \frac{p_n-1}{2}(j'_z-j_z)}
\\
&=-(c_n)_{\boldsymbol j}i (\Gamma/{2})\sum_{\boldsymbol g}\frac{\left[k_0^2-|({\boldsymbol q}_n-{\boldsymbol g})\cdot {\boldsymbol p}|^2 \right]}{k_0\sqrt{k_0^2-|{\boldsymbol q}_n-{\boldsymbol g}|^2}}\left(1+p_n e^{i \sqrt{k_0^2-|{\boldsymbol q}_n-{\boldsymbol g}|^2}L}\right),
\end{aligned}\end{equation}
\end{widetext}
where $\Gamma=3\pi\gamma_e/(k_0\delta_\perp)^2$.

From Eq.~\eqref{eq:eigHexplicit}, a finite number of diffraction orders (i.e., vectors ${\boldsymbol g}$) contribute to the decay rate $\gamma_n$, as for $|{\boldsymbol q}_n-{\boldsymbol g}|\geq k_0$ the eigenvalue becomes purely real. We thus obtain
\begin{equation}
\begin{aligned}\label{eq:gammanflatarrays}
\gamma_n=\Gamma\overline{\sum_{\boldsymbol g}}&\frac{\left[k_0^2-|({\boldsymbol q}_n-{\boldsymbol g})\cdot {\boldsymbol p}|^2 \right]}{k_0\sqrt{k_0^2-|{\boldsymbol q}_n-{\boldsymbol g}|^2}}\\ &\left[1+p_n \cos\left(\sqrt{k_0^2-|{\boldsymbol q}_n-{\boldsymbol g}|^2}L\right)\right]
\end{aligned}
\end{equation}
where $\overline{\sum_{\boldsymbol g}}$ is restricted to vectors ${\boldsymbol g}$ of the reciprocal lattice satisfying $|{\boldsymbol q}_n-{\boldsymbol g}|< k_0$. In particular, for ${\boldsymbol q}_n={\boldsymbol 0}$, a single order ($m_x=m_y=0$) contributes, provided $\delta_\perp<\lambda_0$, in which case we obtain Eq.~\eqref{eq:gammanplanewaves}. Moreover, this becomes valid for all ${\boldsymbol q}_n$ if $\delta_\perp<\lambda_0/2$.

The self-energies $\Delta_n$ can be similarly evaluated, however with a bit of caution. Indeed the real part in Eq.~\eqref{eq:eigHexplicit} diverges as all ${\boldsymbol g}$ with $|{\boldsymbol q}_n-{\boldsymbol g}|\geq k_0$ now contribute. We distinguish between two contributions, i.e. write $\sum_{\boldsymbol g}=\overline{\sum_{\boldsymbol g}}+\overline{\overline{\sum_{\boldsymbol g}}}$, where the second sum accounts for these vectors. Similarly, we write $\Delta_n=\overline{\Delta_n}+\overline{\overline{\Delta_n}}$ with
\begin{equation*}\begin{aligned}
\overline{\Delta_n}=p_n(\Gamma/{2})\sum_{\boldsymbol g}&\frac{\left[k_0^2-|({\boldsymbol q}_n-{\boldsymbol g})\cdot {\boldsymbol p}|^2 \right]}{k_0\sqrt{k_0^2-|{\boldsymbol q}_n-{\boldsymbol g}|^2}}
\\ & \sin\left(\sqrt{k_0^2-|{\boldsymbol q}_n-{\boldsymbol g}|^2}L\right),
\end{aligned}\end{equation*}
while $\overline{\overline{\Delta_n}}$ is obtained numerically from the eigenvalues of $\mathcal H$.  
This last term is independent of $L$ as the exponential term in Eq.~\eqref{eq:eigHexplicit} vanishes and is thus identical to the self-energy of a single 2D array \cite{Shahmoon2017}.

\subsection{Finite-sized (curved) arrays}
We now consider the case of finite atomic arrays. As discussed in the main text, in order to mitigate the spreading of wavepackets for photons propagating between the arrays, we assume the atoms in each array are located along the phase profile of a Gaussian mode $\mathcal E({\boldsymbol r})$. In the following we provide details on this Gaussian mode, derive analytical expressions for the spectrum of $\mathcal H$, and provide in the end a numerical study of the eigenstate distribution.   

\subsubsection{Definition of the Hermite-Gaussian modes and array curvature}
Here we summarize the properties and notations of the Hermite-Gauss modes,
which are solutions of the paraxial equation for light
\mbox{$[\partial_{z}-({i}/{2k_{0}})\nabla_{\perp}^{2}]\text{TEM}_{j,k}({\boldsymbol r})=0$} for modes propagating along $z$,
with \mbox{$\nabla_{\perp}^{2}\equiv\partial_{x}^{2}+\partial_{y}^{2}$}. These modes represent a natural basis for the field generated by the atomic arrays.
Assuming the focal point is here located at ${\boldsymbol r}={\boldsymbol 0}$, these modes are defined
by the waist $w_{0}$ as \cite{hecht}
\begin{equation}
\begin{aligned}
\text{TEM}_{j,k}({\boldsymbol r})=&\sqrt{\tfrac{2}{\pi w(z)^2}}H_i\left(\sqrt{2}x/w(z)\right)H_j\left(\sqrt{2}y/w(z)\right)
\\ &e^{-(x^2+y^2)/w(z)^{2}}e^{i\left(k_{0}(x^2+y^2)/[2R(z)]-\psi_{j,k}(z)\right)},\label{eq:upldef}
\end{aligned}
\end{equation}
with $j,k=(0,1,...)$, $H_j$ is the Hermite polynomial of order $j$, 
\begin{equation}\label{eq:defwidth}
w(z)=w_{0}\sqrt{1+(z/z_{R})^{2}}
\end{equation}
 the mode width, %
\begin{equation}\label{eq:Rzdef}
R(z)=z\left[1+(z_{R}/z)^{2}\right]%
\end{equation} 
the radius of curvature, %
\begin{equation*}
\psi_{j,k}(z)=(j+k+1)\text{tan}^{-1}(z/zR)%
\end{equation*} the Gouy phase and %
$z_{R}=\pi w_{0}^{2}/\lambda_{0}$%
the Rayleigh length, and are normalized as
\begin{equation*}
\int d{\boldsymbol r}_\perp\text{TEM}_{j,k}({\boldsymbol r})(\text{TEM}_{j',k'}({\boldsymbol r}))^*=\delta_{j,j'}\delta_{k,k'}.
\end{equation*}

In particular, the phase profile of the Gaussian mode, which determines the curvature of the atomic arrays, is taken as \mbox{$\mathcal E({\boldsymbol r})\sim \text{TEM}_{0,0}({\boldsymbol r})e^{ik_0 z}$}. Specifically, for a given separation distance $L$ between arrays and mode waist $w_0$, the longitudinal position $z_{\boldsymbol j}$ of atom $\boldsymbol j$ satisfies
\begin{equation}\label{eq:phaseprofile}
k_{0}z_{\boldsymbol j}+k_{0}(x_{\boldsymbol j}^2+y_{\boldsymbol j}^2)/[2R(z_{\boldsymbol j})]-\psi_{0,0}(z_{\boldsymbol j})=\pm k_{0}L/2,
\end{equation}
where the $+$ and $-$ signs correspond respectively to atoms in
the second array ($j_z=2$) and in the first array ($j_z=1$), such that the phase of $\mathcal E({\boldsymbol r}_{\boldsymbol j})$ only depends on $j_z$. From Eq.~\eqref{eq:Rzdef}, the curvature radius for the arrays is maximal when $L\approx 2z_R$, yielding a displacement along $z$ of $\sim L_\perp^2/(2L)$ at the corners of the arrays, which can take values of the order of the wavelength $\lambda_0$. As we show in the next section, the condition of Eq.~\eqref{eq:phaseprofile} allows us to construct non-local eigenstates of $\mathcal H$ with Gaussian distribution.

\subsubsection{Analytical expressions}
We now derive expressions for the eigenvalues of $\mathcal H$. Due to the finite array size, plane waves are not longer eigenstates of $\mathcal H$, however parity remains a symmetry of the system. We thus write the eigenstates as $(c_n)_{({\boldsymbol j}_\perp,1)}=p_n(c_n)_{({\boldsymbol j}_\perp,2)}\equiv (v_n)_{{\boldsymbol j}_\perp}/\sqrt{2}$ with $p_n=\pm1$. The matrix $\mathcal H$ can then be decomposed into 2 matrices of size $N$, namely 
\begin{equation}\label{eq:defH0}
(\mathcal H_0)_{{\boldsymbol j}_\perp,{\boldsymbol j}'_\perp}\equiv (\mathcal H)_{({\boldsymbol j}_\perp,1),({\boldsymbol j}'_\perp,1)}=(\mathcal H)_{({\boldsymbol j}_\perp,2),({\boldsymbol j}'_\perp,2)}
\end{equation}
accounting for the dipole-dipole interaction within each array, and
\begin{equation}\label{eq:defH1}
(\mathcal  H_1)_{{\boldsymbol j}_\perp,{\boldsymbol j}'_\perp}\equiv (\mathcal H)_{({\boldsymbol j}_\perp,1),({\boldsymbol j}'_\perp,2)}=(\mathcal H)_{({\boldsymbol j}_\perp,2),({\boldsymbol j}'_\perp,1)},
\end{equation}
accounting for the effective interaction between different arrays, with $v_n$ being an eigenstate of $\mathcal H_0 + p_n \mathcal H_1$, with the same eigenvalue $\epsilon_n$. 

The Green's tensor in Eq.~\eqref{mathcalH} can be formally decomposed as 
\begin{equation}\label{eq:decompG}
{G}({\boldsymbol r})=\frac{3\pi}{k_{0}^{2}}G_\text{par}({\boldsymbol r}) + G'({\boldsymbol r}),
\end{equation}
where $G_\text{par}({\boldsymbol r})$ is the Green's function for paraxial modes, reading 
\begin{equation}\label{eq:defGpar}G_\text{par}({\boldsymbol r})=\frac{k_0}{2\pi i |z|}e^{ik_0\left[\right|z|+|{\boldsymbol r}_\perp|^2/(2|z|)]},
\end{equation} 
with 
\begin{equation}\label{eq:pargreen}
\int d{\boldsymbol r}'_\perp G_\text{par}({\boldsymbol r}-{\boldsymbol r}')\text{TEM}_{j,k}({\boldsymbol r}')e^{ik_0z'}=\text{TEM}_{j,k}({\boldsymbol r})e^{ik_0z}
\end{equation} 
for $z'>z$.

We define the Fourier transform 
\begin{equation*}
(\tilde v_n)_{\boldsymbol q}=\frac1{\sqrt{N}}\sum_{{\boldsymbol j}_\perp} (v_n)_{{\boldsymbol j}_\perp} e^{i\delta_\perp{\boldsymbol j}_\perp\cdot{\boldsymbol q}},
\end{equation*}
and the mean absolute quasi-momentum 
\begin{equation*}
\overline q=\sum_{\boldsymbol q} |(\tilde v_n)_{\boldsymbol q}|^2 |{\boldsymbol q}|,
\end{equation*}
where $q_x,q_y=-\pi/\delta_\perp+2\pi n_{x,y}/L_\perp$, with $n_{x,y}=0,1,...,N_\perp-1$. We consider in particular eigenstates with low quasi-momentum $\overline {q}$. As we saw in Sec.~\ref{sec:infinitearrays}, provided $\delta_\perp<\lambda_0$ only a single diffraction order contributes to the spontaneous emission, meaning that photons are emitted mostly in the direction normal to the arrays, and as such can be treated within a paraxial approximation. This motivates us to look for eigenvectors distributed as $(v_{(j,k)})_{{\boldsymbol j}_\perp}\sim \text{TEM}_{j,k}({\boldsymbol r}_{({\boldsymbol j}_\perp,1)})e^{i k_0 z_{({\boldsymbol j}_\perp,1)}}$. 
From Eqs.~\eqref{eq:ME2nodes}, \eqref{eq:decompG} and \eqref{eq:pargreen}, we then have
\begin{equation*}
\begin{aligned}
\sum_{{\boldsymbol j}'_\perp}&(\mathcal H_0)_{{\boldsymbol j}_\perp,{\boldsymbol j}'_\perp} (v_{(j,k)})_{{\boldsymbol j}'_\perp} 
\\&\approx \left(\overline{\overline{\Delta}}_{(j,k)} -i\Gamma/2 -i \gamma'_{(j,k)}/2\right) (v_{(j,k)})_{{\boldsymbol j}'_\perp},
\end{aligned}
\end{equation*}
where $\Gamma=3\pi\gamma_e/(k_0\delta_\perp)^2$, and we approximated the sum as an integral $\sum_{{\boldsymbol j}'_\perp}\approx \int d{\boldsymbol r}'_\perp/\delta_\perp^2$. The term $\gamma'_{(j,k)}$ is a phenomenological decay by photon emission into non-paraxial modes, which we add as a perturbation accounting for $G'({\boldsymbol r})$ in Eq.~\eqref{eq:decompG}. In analogy to Sec.~\ref{sec:infinitearrays}, the self-energy $\overline{\overline{\Delta}}_{(j,k)}$ on the other hand diverges due to the divergence in Eq.~\eqref{eq:defGpar} when $z\to 0$, and must be evaluated numerically. $v_{(j,k)}$ is thus approximately eigenstate of $\mathcal H_0$.
Similarly, we get 
\begin{equation*}
\sum_{{\boldsymbol j}'_\perp}(\mathcal H_1)_{{\boldsymbol j}_\perp,{\boldsymbol j}'_\perp} (v_{(j,k)})_{{\boldsymbol j}'_\perp} \approx -i(\Gamma/2) \text{TEM}_{j,k}({\boldsymbol r}_{({\boldsymbol j}_\perp,2)})e^{i k_0 z_{({\boldsymbol j}_\perp,2)}},
\end{equation*}
where the effect of $G'({\boldsymbol r})$ is here neglected in a paraxial approximation for the photons exchanged between different arrays.

\begin{figure}
\includegraphics[width=0.5\textwidth]{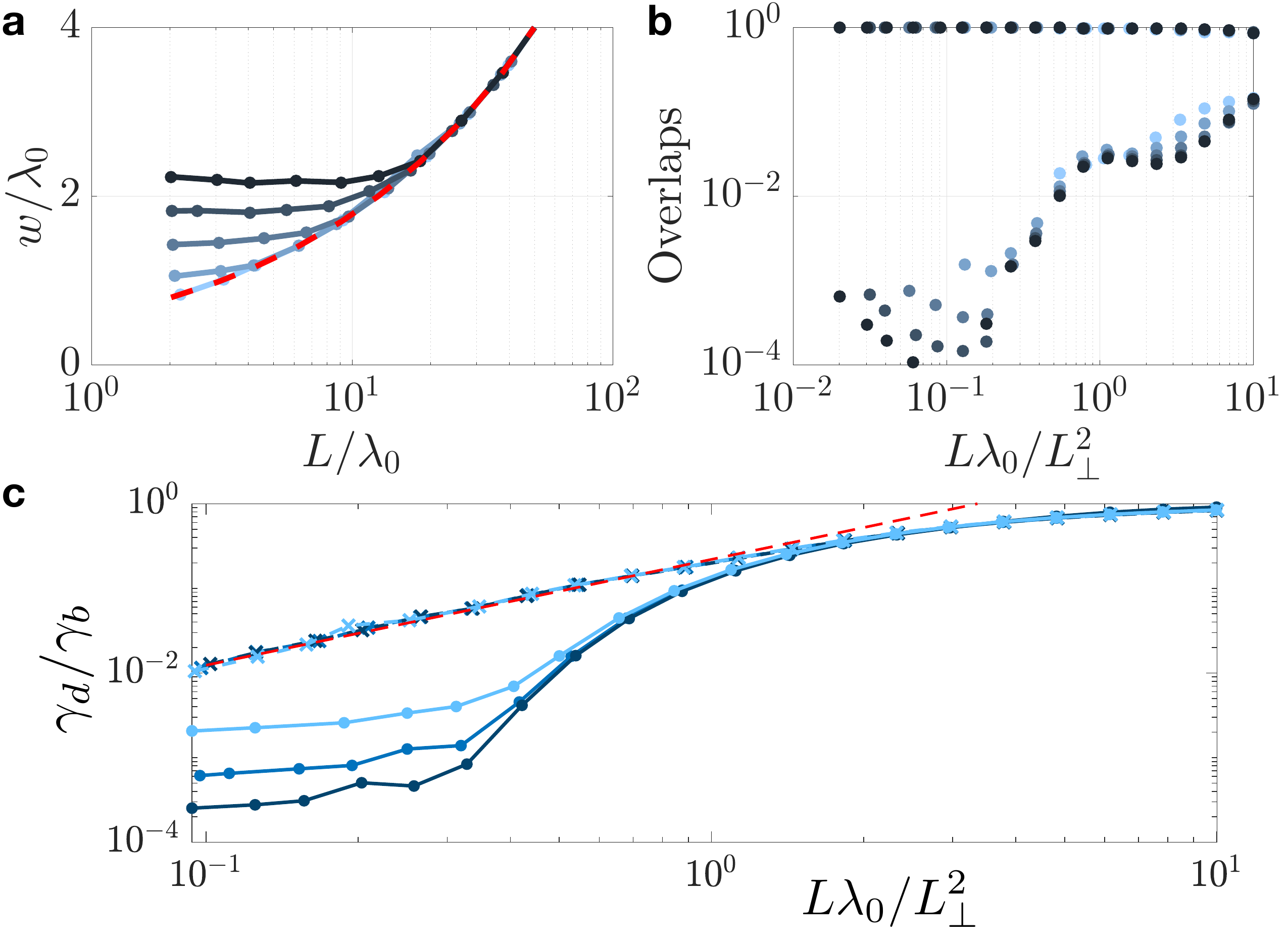}
\caption{\label{fig:A2} (a)~Optimal mode width $w(z=L/2)$, with $\sqrt{L/(\pi \lambda_0)}$ in dashed red. (b)~Overlaps $O_n$ of eigenstates with Gaussian distribution, for dark and bright states (upper points) and all other states (lower points). $\delta_\perp=\lambda_0/2$, $N_\perp=4,8,12,16,20$ (light to dark blue). (c)~Ratio of dark and bright state decay rates with $N_\perp=8,12,16$ (light to dark blue) and $\delta_\perp=0.5\lambda_0$, for curved (dots) and flat (crosses) arrays. Dashed red: $\sim (L\lambda_0/L_\perp^2)^{1.25}$.}
\end{figure}

Using Eq.~\eqref{eq:phaseprofile}, we have 
\begin{equation*}
\text{TEM}_{0,0}({\boldsymbol r}_{({\boldsymbol j}_\perp,2)})e^{i k_0 z_{({\boldsymbol j}_\perp,2)}}=e^{ik_0L}(v_{(0,0)})_{{\boldsymbol j}_\perp},
\end{equation*} 
such that $v_{(0,0)}$ is eigenstate of $\mathcal H_0 \pm \mathcal H_1$ with eigenvalue
\begin{equation}\label{eq:epsilon00}
\epsilon_{(0,0)}=\overline{\overline{\Delta}}_{0,0} -i\Gamma/2 \left(1\pm e^{ik_0L}\right) -i \gamma'_{(0,0)}/2.
\end{equation} 
This is the expression of the eigenvalues for the symmetric and anti-symmetric eigenstates of the main text, where we identify $\gamma_d\equiv\gamma'_{(0,0)}$ and $\Delta_d\equiv\overline{\overline{\Delta}}_{0,0}$. For $k_0L=m\pi$ with integer $m$ one of these states is thus `dark' (with minimal decay $\gamma_d$), while the other state is `bright' (as it decays with rate $\gamma_b=2\Gamma+\gamma_d$).
For $(j,k)\neq(0,0)$ on the other hand, we get similar expressions by considering that the Gouy phase $\psi_{j,k}$ in Eq.~\eqref{eq:upldef} is approximately constant for atoms within the same array. We then have 
\begin{equation*}
\text{TEM}_{j,k}({\boldsymbol r}_{({\boldsymbol j}_\perp,2)})e^{i k_0 z_{({\boldsymbol j}_\perp,2)}}=e^{ik_0L}e^{i\phi_{j,k}}(v_{(j,k)})_{{\boldsymbol j}_\perp},
\end{equation*} 
where 
\begin{equation}\label{eq:exprphijk}
\phi_{j,k}=2(j+k)\text{tan}^{-1}[L/(2z_R)],
\end{equation} such that $v_{(j,k)}$ is eigenstate of $\mathcal H_0 \pm \mathcal H_1$ with eigenvalue
\begin{equation}\label{eq:epsilonTEMjk}
\epsilon_{(j,k)}=\overline{\overline{\Delta}}_{j,k} -i\Gamma/2 \left(1\pm e^{ik_0L}e^{i\phi_{j,k}}\right) -i \gamma'_{(j,k)}/2.
\end{equation}

\begin{figure*}
\includegraphics[width=0.85\textwidth]{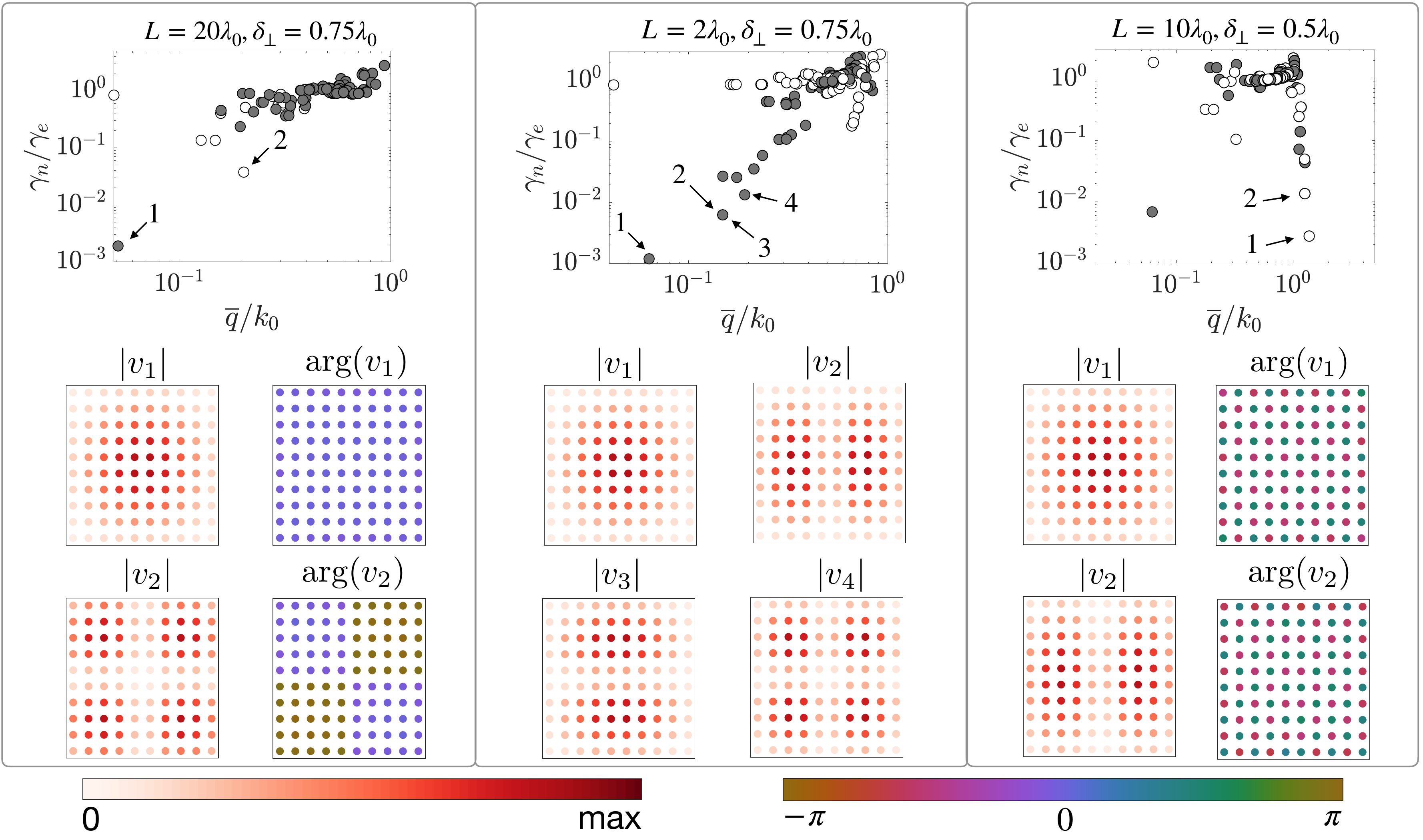}
\caption{\label{fig:Hanalysis}\emph{Spectrum of $\mathcal H$.}  First row: decay rates $\gamma_n$ and average quasi-momentum $\bar q$ of the eigenstates of $\mathcal H$, with $N_\perp=10$. An even (odd) parity is denoted with a white (resp. black) circle. Second and third rows: distribution of probability amplitude of eigenstates $|v_n|$ in each 2D array, and of phase arg$(v_n)$. }
\end{figure*}

\subsubsection{Numerical study}
Here we provide details on the eigenstates $v_n$ and eigenvalues $\epsilon_n$ of $\mathcal H$ for finite arrays. The decay rates $\gamma_d$ and $\gamma_b$ are obtained by diagonalizing $\mathcal H$ and identifying the dark and bright states as the eigenstates with lowest quasi-momentum $\overline q$. We minimize the ratio $\gamma_d/\gamma_b$ by varying $w_0$, which sets the longitudinal atomic according to Eq.~\eqref{eq:phaseprofile}.

In Fig.~\ref{fig:A2}(a) we show the optimal mode width $w$ [from Eq.~\eqref{eq:defwidth}] for the parameters of Fig.~\ref{fig:fig2}(a) for a single excitation. At large $L \gtrsim L_\perp^2/\lambda_0$, we have $w=\sqrt{L\lambda_0/\pi}$, which is the minimal width achievable for fixed $L$ within the diffraction limit, where $z_R=L/2$. In this regime imperfections (i.e., finite $\gamma_d$)  are mainly due to the array size being too small to fit a Gaussian mode connecting the arrays. At small $L\ll  L_\perp^2/\lambda_0$, the width saturates to around $w\sim L_\perp/4$. This is a trade-off between having the Gaussian mode $\mathcal E({\boldsymbol r})$ fit the arrays, and increasing the number of participating atoms in order to minimize the emission to non-paraxial modes, both effects leading to a finite rate $\gamma_d$. In Fig.~\ref{fig:A2}(b) we represent the overlap of the eigenvectors $v_n$ with the Gaussian mode $\mathcal E({\boldsymbol r})$. For each eigenvectors, this overlap is computed as $O_n=\left|\sum_{{\boldsymbol j}_\perp} \mathcal E({\boldsymbol r}_{({\boldsymbol j}_\perp,1)}) (v_n)^*_{{\boldsymbol j}_\perp}\right|^2/\sum_{{\boldsymbol j}_\perp}|\mathcal E({\boldsymbol r}_{({\boldsymbol j}_\perp,1)})|^2$. The overlaps for the dark and bright modes are represented as the upper points, and are close to $1$. Conversely, the sum of the overlaps of all other eigenstates is represented as the lower points, and takes very small values, vanishing for $L\lesssim L_\perp^2/\lambda_0$. 

The requirement for curving the atomic arrays is studied in Fig.~\ref{fig:A2}(c), where we compare the ratio of $\gamma_d/\gamma_b$ between flat arrays (for which $z_{\boldsymbol j}=\pm L/2$) and curved arrays (satisfying Eq.~\eqref{eq:phaseprofile}). One sees that the curvature can improve this ratio by several orders of magnitude for $L\lesssim L_\perp^2/\lambda_0$.

In Fig.~\ref{fig:Hanalysis} we show the decay rates of all eigenstates $\gamma_n$ as well as their average quasi-momenta $\overline q$ for $N_\perp=10$. On the left, we represent the same situation as in Fig.~\ref{fig:setup}. We note first that the dark state, labeled $1$, as well as the bright state above, have the distribution of a $\text{TEM}_{0,0}$ mode. The second most subradiant state, labeled $2$, corresponds to a $\text{TEM}_{1,1}$ mode. We note that its parity is opposite to that of $v_1$, which is due to the fact that here $L$ is large enough that $z_R=L/2$, and $\phi_{1,1}=\pi$ in Eq.~\eqref{eq:exprphijk}. On the other hand, eigenvectors distributed according to $\text{TEM}_{1,0}$ and $\text{TEM}_{0,1}$ modes cannot be subradiant as they are out of phase, with $\phi_{1,0}=\phi_{0,1}=\pi/2$.

In the second column, we consider the situation with $L=2\lambda_0$, which corresponds to the opposite extreme regime where $z_R\gg L$, and $\phi_{j,k}\approx 0$, such that now the subradiant states have the same parity, and eigenvectors with $\text{TEM}_{1,0}$ and $\text{TEM}_{0,1}$ distributions can also be subradiant. The decay rate significantly increases with $\overline q$, which can be understood as a gradual breaking of the paraxial approximation, and can be seen from Eq.~\eqref{eq:gammanflatarrays} as the interference between arrays becomes imperfect.

Finally, in the third column we show that subradiance can also appear in eigenvectors with large $\overline q$ when $\delta_\perp<\lambda_0/\sqrt{2}$. There, subradiance is due to the fact that these guided modes have their momentum larger than $k_0$, and as such reside outside of the light cone, and are studied e.g. in Refs.~\cite{Sutherland2016,Asenjo-Garcia2017}. These modes are localized in each array, and are thus degenerate in energy, in contrast to the non-local modes with low $\overline q$.

\subsection{Dark and bright state probing}\label{sec:probing}
\begin{figure}
\includegraphics[width=1\columnwidth]{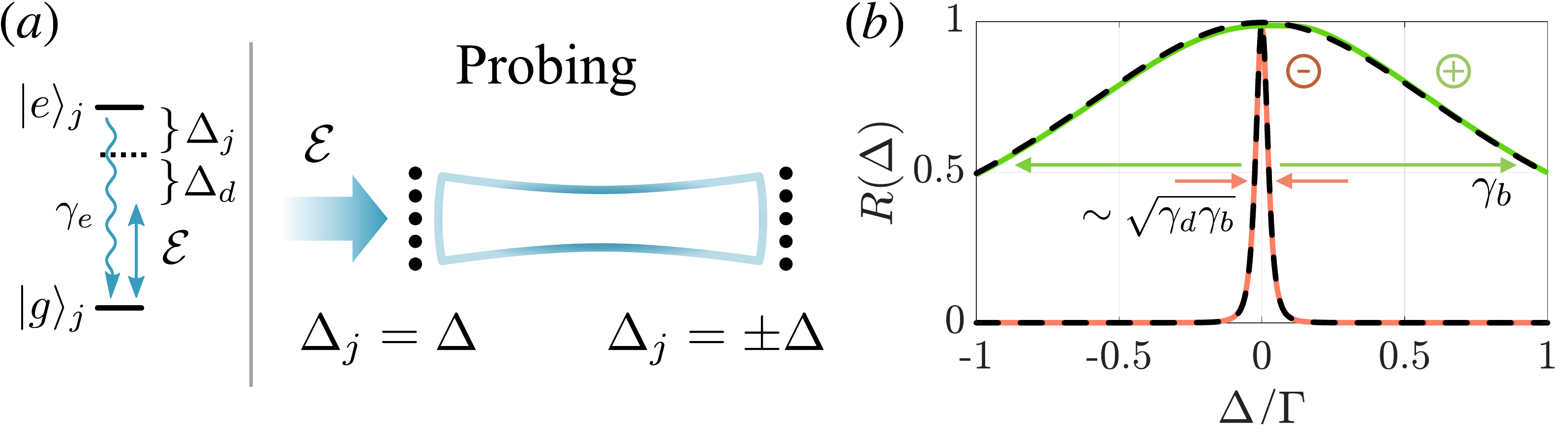} 
\caption{\textit{Dark and bright state probing.} (a)~Level scheme and sketch for probing the dark and bright states. A weak probe field $\mathcal E$, resonant with the collective atomic shift $\Delta_d$ and propagating in the Gaussian mode $\mathcal E({\boldsymbol r})$, drives the atomic arrays, while the transition frequency of the atoms in the first and second arrays is shifted by $\Delta$, either symmetrically or with opposite sign. (b)~The reflection probability displays a peak with the width given by $\gamma_b$ (for symmetric shift, in green) or $\sqrt{\gamma_d\gamma_b}$ (for opposite shift, in red), with $N_\perp=12$, $L=30\lambda_0$, $\delta_\perp=0.8\lambda_0$. Dashed black: analytical expressions.
\label{fig:fig4}}
\end{figure}

Finally, here we show how to probe the dark and bright state lifetimes in the reflectivity of a laser. The situation is represented in Fig.~\ref{fig:fig4}(a).
We consider a weak laser with polarization ${\boldsymbol p}$ propagating along $z$ in the Gaussian mode $\mathcal E({\boldsymbol r})$ at frequency $\omega_0+\Delta_d$, and driving the system with the atoms in their ground state $\ket{\mathcal G}=\otimes_{\boldsymbol j}\ket{g}_{\boldsymbol j}$. Assuming the field is weak enough, each photon will be scattered by the system independently, and the reflectivity can be evaluated for single-photon pulses. We can thus write the state of the system for a single excitation as 
\begin{equation*}\label{eq:eq2mt}
\ket{\psi(t)}=\sum_{{\boldsymbol j}}c_{\boldsymbol j}(t)\sigma_{\boldsymbol j}^+\ket{\mathcal G}\ket{0}+\int d{\boldsymbol k}\sum_\lambda \psi_\lambda({\boldsymbol k},t)\ket{\mathcal G}\ket{{\boldsymbol k},\lambda},
\end{equation*}  
 with here $c_{\boldsymbol j}(0)=0$. We assume moreover that atoms in each array $j_z$ additionally detuned by $\Delta_{j_z}$. We consider two situations, with either $\Delta_{j_z}=\Delta$ (i.e., with the same detuning for atoms in both arrays), or $\Delta_z=2\Delta(3/2-j_z)$ (with opposite detuning between the two arrays). 
 
 As in Sec.~\ref{sec:opticalmodel}, the field dynamics can be integrated, yielding for the atoms 
 \begin{equation}\label{eq:dotcj}
 \begin{aligned}
 \dot c_{\boldsymbol j}=&-i\sum_{\boldsymbol j'}\left(\mathcal H_{\boldsymbol j,\boldsymbol j'}+(\Delta_{j_z}-\Delta_d)\delta_{\boldsymbol j,\boldsymbol j'}\right) c_{\boldsymbol j'}(t)
 \\ &- \sqrt{\frac{3\pi c \gamma_e}{2k_0^2}} \psi^\text{in}({\boldsymbol r}_{\boldsymbol j},t),
 \end{aligned}
 \end{equation}
where we moved to a frame rotating with $\omega_0+\Delta_d$, and defined the input field
\begin{equation*}
\begin{aligned}
 \psi^\text{in}({\boldsymbol r},t)=\frac1{\sqrt{(2\pi)^3}}\int d{\boldsymbol k}&e^{i {\boldsymbol k}\cdot {\boldsymbol r}}e^{-i(\omega_{\boldsymbol k}-\omega_0-\Delta_d)t}
 \\ & \sum_{\lambda}{\boldsymbol p}^*\cdot {\boldsymbol e}_{\lambda,{\boldsymbol k}}\psi_{\lambda}({\boldsymbol k},0).
\end{aligned}
\end{equation*}
The resulting field on the other hand reads, neglecting retardation effects,
\begin{equation}\label{eq:psirt1}
\psi({\boldsymbol r},t)=\psi^\text{in}({\boldsymbol r},t) -i\sqrt{\frac{\gamma_ek_0^2}{6\pi c}} \sum_{\boldsymbol j}c_{\boldsymbol j}(t){G}({\boldsymbol r}-{\boldsymbol r}_{\boldsymbol j})
\end{equation}
For long pulses, i.e., varying over timescales much larger than the atomic response time $1/\gamma_d$, we can set $\dot c_{\boldsymbol j}\approx0$ in Eq.~\eqref{eq:dotcj}, and get from Eq.~\eqref{eq:psirt1}
\begin{equation}\label{eq:psirt2}
\begin{aligned}
\psi({\boldsymbol r},t)=\psi^\text{in}({\boldsymbol r},t) -&\frac{\gamma_e}{2} \sum_{\boldsymbol j, \boldsymbol j'}{G}({\boldsymbol r}-{\boldsymbol r}_{\boldsymbol j})
\\ &\left(\mathcal H+ \mathbb \Delta_{z}-\Delta_d\mathbb 1\right)^{-1}_{\boldsymbol j,\boldsymbol j'}\psi^\text{in}({\boldsymbol r}_{\boldsymbol j'},t),
\end{aligned}
\end{equation}
where $(\mathbb \Delta_{z})_{\boldsymbol j,\boldsymbol j'}=\Delta_{j_z}\delta_{\boldsymbol j,\boldsymbol j'}$.

The reflectivity $R$ is then obtained by taking $\psi^\text{in}({\boldsymbol r},t)=\mathcal E({\boldsymbol r})$ as the overlap between $\psi({\boldsymbol r})$ and a target mode $\psi^\text{tar}({\boldsymbol r})$
\begin{equation*}
R=\left| \int d{\boldsymbol r}_\perp (\psi^\text{tar}({\boldsymbol r}))^*\psi({\boldsymbol r})\right|^2,
\end{equation*}
where the target mode is the Gaussian mode propagating to the left, i.e., $\psi^\text{tar}({\boldsymbol r})=(\mathcal E({\boldsymbol r}))^*$. Within a paraxial approximation for the Gaussian mode we replace the Green's function by its paraxial counterpart in Eq.~\eqref{eq:decompG}, and apply Eq.~\eqref{eq:pargreen} to obtain
\begin{equation}\label{eq:Rnum}
R=\frac{9\pi^2\gamma_e^2}{4k_0^4}\left|\sum_{\boldsymbol j,\boldsymbol j'} \mathcal E({\boldsymbol r}_{\boldsymbol j})\left(\mathcal H+ \mathbb \Delta_{z}-\Delta_d\mathbb 1\right)^{-1}_{\boldsymbol j,\boldsymbol j'}  \mathcal E({\boldsymbol r}_{\boldsymbol j'}) \right|^2
\end{equation}
which can be evaluated numerically, and provides the green and red curves in Fig.~\ref{fig:fig4}(b). 
We stress that the expression of Eq.~\eqref{eq:Rnum} is valid only within the paraxial approximation. This breaks down for configurations with $w_0\lesssim\lambda_0$, which can occur for small arrays with $L_\perp\lesssim 4\lambda_0$. 

As we saw from Fig.~\ref{fig:A2}(b), only the dark and bright states have significant overlap with the Gaussian distribution $\mathcal E({\boldsymbol r}_{\boldsymbol j})$, such that we can restrict the vector space of the matrix $\mathcal H+ \mathbb \Delta_{z}-\Delta_d\mathbb 1$ to these two states, and invert it on this subspace. Approximating $\sum_{\boldsymbol j}\approx \sum_{j_z}\int d{\boldsymbol r}_{\perp}/\delta_\perp^2$ and using Eq.~\eqref{eq:phaseprofile}, we obtain $R(\Delta)= (\gamma_b-\gamma_d)^2/(\gamma_b^2+4\Delta^2)$ for the case of symmetric detuning between the two arrays, and $R(\Delta) = (\gamma_b-\gamma_d)^2/(\gamma_b+4\Delta^2/\gamma_d)^{2}$ for the case of opposite detuning, with $R=1$ if $\Delta=0$. These expressions are represented in the dashed black curves of Fig.~\ref{fig:fig4}(b), which show excellent agreement with the numerical results.

\section{Photonic link between quantum memories}
Here we write an effective model for the atomic dynamics, retaining four modes as expressed in Eq.~\eqref{eq:Heff}.
We then derive the expression for the fidelity of quantum state transfer in Eq.~\eqref{eq:Fopt}, and explain how to write and read from the local quantum memory states. We finally discuss how our results extend to non-markovian regimes, where retardation effects due to the finite speed of photons exchanged between arrays is no longer negligible.

\subsection{Effective four mode model}\label{sec:darkbrightmodes}
We now consider each atom has a $\Lambda$ level structure, as represented in Fig.~\ref{fig:fig3}(a). We assume at most a single atom is in state $\ket{e}$ or $\ket{s}$ at a time, and a laser drives the $\ket{e}\to\ket{s}$ transition resonantly with the cooperative shift $\Delta_d$, with homogeneous Rabi frequency $\Omega$. In a rotating frame, the system is thus described by the master equation in Eq.~\eqref{eq:ME2nodes}, with
\begin{equation*}
H_\text{drive}=-\Delta_d\sum_{\boldsymbol j}\sigma_{\boldsymbol j}^+\sigma_{\boldsymbol j}^-+\Omega\sum_{\boldsymbol j}(s_{\boldsymbol j}^+\sigma_{\boldsymbol j}^-+\text{h.c.}),
\end{equation*}
with $s_{\boldsymbol j}^+=\ket{s}_{\boldsymbol j}\!\bra{g}$. We start from an initial state of the form 
\begin{equation}\label{eq:psiidef}
\ket{\psi_i}=c_g\ket{\mathcal G}+c_sS_1^+\ket{\mathcal G},
\end{equation}
where $S_1^+=\sum_{{\boldsymbol j}_\perp}(v_d)_{{\boldsymbol j}_\perp} s_{(\boldsymbol j_\perp,1)}^+$ creates an excitation of $\ket{s}$ in the first array, with $(v_d)_{\boldsymbol j_\perp}\sim \mathcal E({\boldsymbol r}_{(\boldsymbol j_\perp,1)})$ the probability amplitude corresponding to the dark (and bright) state. Similarly we define $S_2^+=\sum_{{\boldsymbol j}_\perp}(v_d)_{{\boldsymbol j}_\perp} s_{(\boldsymbol j_\perp,2)}^+$, which creates an excitation of $\ket{s}$ in the second array. The dynamics of Eq.~\eqref{eq:ME2nodes} will then excite only the eigenstates of $\mathcal H$ with Gaussian distribution, i.e., the dark and bright states, as shown in Fig.~\ref{fig:A2}(b). We thus obtain an effective model with only four modes: the two local modes, with creation operators $S_1^+$ and $S_2^+$, as well as the dark and bright modes, created by the operators
\begin{equation*}
\sigma^+_{b/d}=\sum_{\boldsymbol j_\perp}(v_d)_{\boldsymbol j_\perp}\left(\sigma^+_{(\boldsymbol j_\perp,1)}\pm (-1)^m\sigma^+_{(\boldsymbol j_\perp,2)} \right)/\sqrt{2}.
\end{equation*}
The atomic dynamics thus follows
\begin{equation}\label{eq:mastereqheff}
\frac{d\rho}{dt}=-i[H_\text{eff},\rho]+\gamma_b\mathcal D[\sigma_b^-]\rho+\gamma_d\mathcal D[\sigma_d^-]\rho,
\end{equation}
with 
\begin{equation*}
H_\text{eff}=\frac{\Omega}{\sqrt{2}}\left[ \sigma_b^+\left(S_1^-+S_2^-\right)+\sigma_d^+\left(S_1^--S_2^-\right) \right]+\text{h.c.}
\end{equation*}
and $\mathcal D[a]\rho = a\rho a^\dagger -(1/2)(a^\dagger a\rho + \rho a^\dagger a)$. We note that Fig.~\ref{fig:fig3} provides a numerical verification of this four modes model for the simulation of quantum state transfer, which we describe below.

\subsection{Quantum state transfer}\label{sec:solutioneffectivemodel}
We now provide an analytical derivation of the fidelity for quantum state transfer. Assuming the system is initially prepared in the pure state $\ket{\psi_i}$ of Eq.~\eqref{eq:psiidef}, the atomic density matrix can be expressed as 
\begin{equation*}
\rho(t)=\ket{\psi(t)}\bra{\psi(t)} + P_g(t)\ket{\mathcal G}\bra{\mathcal G},
\end{equation*}
where 
\begin{equation*}
\left\vert \psi(t)\right\rangle =\left(c_g+c_1(t)s_{1}^{+}+c_2(t)s_{2}^{+}+c_{b}(t)\sigma_{b}^{+}+c_{d}(t)\sigma_{d}^{+}\right)\left\vert {\mathcal G}\right\rangle,
\end{equation*} 
with $c_1(0)=c_s$, $c_2(0)=c_b(0)=c_d(0)=P_g(0)=0$. We wish to transfer the quantum state $\ket{\psi_i}$ to the second array, i.e., have the system evolve to 
\begin{equation*}\label{eq:psifdef}
\ket{\psi_f}=c_g\ket{\mathcal G}+c_sS_2^+\ket{\mathcal G}.
\end{equation*}
We define the fidelity of the state transfer as %
\mbox{%
${\cal F}\equiv\max_{t}\left|c_{2}\left(t\right)\right|^{2}$%
} for $c_s=1$. 
Eq.~\eqref{eq:mastereqheff} then yields
\begin{equation}
\begin{aligned}
\dot c_{1}\left(t\right) & =-i\frac{\Omega}{\sqrt{2}}\left( c_{b}\left(t\right)+c_{d}\left(t\right)\right) \label{eq:s1}\\
\dot c_{2}\left(t\right) & =-i\frac{\Omega}{\sqrt{2}}\left( c_{b}\left(t\right)-c_{d}\left(t\right)\right) \\
\dot c_{b}\left(t\right) & =-\frac{\gamma_{b}}{2}c_{b}\left(t\right)-i\frac{\Omega}{\sqrt{2}}\left( c_{1}\left(t\right)+c_{2}\left(t\right)\right) \\
\dot c_{d}\left(t\right) & =-\frac{\gamma_{d}}{2}c_{d}\left(t\right)-i\frac{\Omega}{\sqrt{2}}\left( c_{1}\left(t\right)-c_{2}\left(t\right)\right).
\end{aligned}
\end{equation}
The general solution of Eq.~\eqref{eq:s1} for $c_{2}\left(t\right)$
can be written in the form 
\begin{equation}
c_{2}\left(t\right)=\sum_{i=\pm1}C_{b,i}e^{-i\omega_{b,i}t}+\sum_{i=\pm1}C_{d,i}e^{-i\omega_{d,i}t},\label{eq:C_2}
\end{equation}
where, in the regime $\gamma_{d}\ll \Omega\ll\gamma_{b}$, 
\begin{equation}
\begin{aligned}
\omega_{d,\pm1} & =\frac{1}{4}\left( -i\gamma_{d}\pm g\right), \label{eq:w_d}\\
\omega_{b,\pm1} & =\frac{1}{4}\left( -i\gamma_{b}\pm\sqrt{16\Omega^{2}-\gamma_{b}^{2}}\right),
\end{aligned}
\end{equation}
with $g\equiv\sqrt{16\Omega^{2}-\gamma_{d}^{2}}$ the frequency
of the oscillations. The corresponding amplitudes can be readily
obtained using the initial conditions, and read
\begin{equation}
\begin{aligned}
C_{d,\pm1} & =\frac{\Omega}{g}e^{\pm i\arctan\left[\gamma_{d}/g\right]}\\
C_{b,\pm1} & =\frac{-4\Omega^{2}}{16\Omega^{2}-\gamma_{b}^{2}\pm\gamma_{b}\sqrt{\gamma_{b}^{2}-16\Omega^{2}}}.\label{eq:C_b}
\end{aligned}
\end{equation}
Using Eqs.~\eqref{eq:C_2}, \eqref{eq:w_d} and \eqref{eq:C_b}, the first maximum
of $c_{2}\left(t\right)$ is approximately
at half the period of Rabi oscillations, i.e. $t_{\text{max}}\approx\left( \pi-\arctan\left[\gamma_{d}/g\right]\right) /g$.
The fidelity $\mathcal F$, given by $c_{2}\left(t_{\text{max}}\right)$, depends on the drive $\Omega$. Expanding
$c_2(t_\text{max})$ up to the first order in $\Omega/\gamma_{b}$ and $\gamma_{d}/\Omega$, we get
\[
c_{2}\left(t_{\text{max}}\right)=-1+\frac{\Omega\pi}{\gamma_{b}}+\frac{\pi\gamma_{d}}{8\Omega} + \mathcal O(\Omega/\gamma_b)^2+\mathcal O(\gamma_d/\Omega)^2.
\]
The optimal drive then reads \mbox{$\Omega_{\text{opt}}=\sqrt{\gamma_{b}\gamma_{d}/8}$}
and the corresponding optimal fidelity of the state transfer is 
\[
{\cal F}_{\text{opt}}\equiv\left|c_{2}\left(t_{\text{max}}\right)\right|^{2}=1-\pi\sqrt{2\gamma_{d}/\gamma_{b}} + \mathcal O(\gamma_d/\gamma_b),
\]
which is Eq.~\eqref{eq:Fopt}.

\subsection{Write and read of quantum memory using single photon pulses}
We now discuss how one can write and read from the quantum memories in the arrays. In particular, assuming the atoms are in state $\ket{\psi_i}$ as in Eq.~\eqref{eq:psiidef} while the photonic field is in the vacuum state $\ket{0}$, we show that the atoms can be brought to their ground state $\ket{\mathcal G}$ while emitting a photonic qubit $c_g\ket{0}+c_s\ket{1}$, where $\ket{1}$ denotes a state with a single photon leaving the system in a well defined spatio-temporal mode, propagating in a given direction. The time-reversed process allows one to absorb a photonic qubit, thereby preparing the atoms in state $\ket{\psi_i}$.

We make the following two additional assumptions. First, the phase acquired by a photon propagating between the arrays $k_0 L$ can be modified, e.g. by slightly changing the distance $L$ over a range of $\sim\lambda_0/2$. Second, the laser drive $\Omega$ can be turned off for the atoms in the second array. For convenience, we consider the system prepared in state $S_1^+\ket{\mathcal G}$, with the laser driving only the first array. The dynamics, and in particular the spatio-temporal shape of the emitted photon, can be obtained following the steps in Sec.~\ref{sec:probing}. We write here the state as 
\begin{equation*}
\begin{aligned}
\ket{\psi(t)}=&\sum_{{\boldsymbol j}}(c_{\boldsymbol j}(t)\sigma_{\boldsymbol j}^++\tilde c_{\boldsymbol j}(t)s_{\boldsymbol j}^+)\ket{\mathcal G}\ket{0}
\\ &+\int d{\boldsymbol k}\sum_\lambda \psi_\lambda({\boldsymbol k},t)\ket{\mathcal G}\ket{{\boldsymbol k},\lambda},
\end{aligned}
\end{equation*}  
with $\tilde c_{\boldsymbol j}(0)=(v_d)_{\boldsymbol j_\perp} \delta_{j_z,1}$ and $c_{\boldsymbol j}(0)=\psi_\lambda({\boldsymbol k},0)=0$. Integrating the field dynamics, we get
\begin{equation}\label{eq:dotcjmemory}
\begin{aligned}
 \dot c_{\boldsymbol j}=&-i\sum_{\boldsymbol j'}\left(\mathcal H_{\boldsymbol j,\boldsymbol j'}-\Delta_d\delta_{\boldsymbol j,\boldsymbol j'}\right) c_{\boldsymbol j'}(t)-i\Omega\delta_{j_z,1}\tilde c_{\boldsymbol j}(t),
 \\ \dot{\tilde c}_{\boldsymbol j}=&-i\Omega \delta_{j_z,1}c_{\boldsymbol j}(t),
 \end{aligned}
 \end{equation}
where we moved to a frame rotating with $\omega_0+\Delta_d$. Assuming $k_0L \neq m\pi$ with integer $m$, neither the symmetric or anti-symmetric Gaussian states are dark, and their decay rate to paraxial modes is given by $\Gamma(1\pm \cos[k_0L])$ [see Eq.~\eqref{eq:epsilon00}]. Thus, provided $\Omega\ll |\Gamma(1\pm \cos[k_0L])|$ (ideally by setting $\cos[k_0L]=0$), the population of state $\ket{e}_{\boldsymbol j}$ can be adiabatically eliminated, i.e., we set $\dot c_{\boldsymbol j}\approx 0$ in Eq.~\eqref{eq:dotcjmemory}, yielding
\begin{equation}\label{eq:dottildecjmemory}
\begin{aligned}
c_{\boldsymbol j}(t)=&-\Omega\sum_{\boldsymbol j'}\left(\mathcal H-\Delta_d\mathbb 1\right)^{-1}_{\boldsymbol j,\boldsymbol j'} \delta_{j'_z,1}\tilde c_{\boldsymbol j'}(t),
 \\ \dot{\tilde c}_{\boldsymbol j}=&i\Omega^2 \delta_{j_z,1}\sum_{\boldsymbol j'}\left(\mathcal H-\Delta_d\mathbb 1\right)^{-1}_{\boldsymbol j,\boldsymbol j'} \delta_{j'_z,1}\tilde c_{\boldsymbol j'}(t).
 \end{aligned}
 \end{equation}
 Next we note that for the initial condition above, from Eq.~\eqref{eq:dottildecjmemory} we have $\tilde c_{\boldsymbol j}(t)=\tilde c(t)(v_d)_{\boldsymbol j_\perp} \delta_{j_z,1}$. Restricting $\mathcal H-\Delta_d\mathbb 1$ to the space spanned by the symmetric and anti-symmetric Gaussian states, we get 
 \begin{equation*}
 \dot{\tilde c}=2\frac{\Omega^2(\Gamma+\gamma_d)}{e^{2ik_0L}\Gamma^2-(\Gamma+\gamma_d)^2}\tilde c(t),
 \end{equation*}
 i.e., the memory will spontaneous emit a photon with rate 
 \begin{equation*}
 \tilde \gamma=-4\text{Re}\left(\frac{\Omega^2(\Gamma+\gamma_d)}{e^{2ik_0L}\Gamma^2-(\Gamma+\gamma_d)^2}\right).
 \end{equation*}
 We remark that $\tilde \gamma$ is independent of $k_0L$ if $\gamma_d\approx0$, and reduces to $\tilde \gamma=2\Omega^2/\Gamma$. 
 The spatio-temporal shape of the outgoing photon is obtained from Eq.~\eqref{eq:psirt1}, as
 \begin{equation*}
\begin{aligned}
\psi({\boldsymbol r},t)=& i\tilde c(t)\Omega\sqrt{\frac{\gamma_ek_0^2}{6\pi c}} 
\\ &\sum_{\boldsymbol j,\boldsymbol j'} {G}({\boldsymbol r}-{\boldsymbol r}_{\boldsymbol j})\left(\mathcal H-\Delta_d\mathbb 1\right)^{-1}_{\boldsymbol j,\boldsymbol j'} \delta_{j'_z,1}(v_d)_{\boldsymbol j_\perp'}.
\end{aligned}
\end{equation*}
We note that the temporal distribution can be tailored by varying $\Omega$ in time. 

The flux of photons emitted in the Gaussian mode $\mathcal E({\boldsymbol r})$ (i.e., propagating to the right) is obtained as the overlap
\begin{equation*}
\begin{aligned}
&P_\rightarrow(t)=c \left| \int d{\boldsymbol r}_\perp \left( \mathcal E({\bf r}) \right)^*\psi({\boldsymbol r},t) \right|^2
\\ =&|\tilde c(t)|^2 \frac{3\pi\gamma_e\Omega^2}{2 k_0^2}\left| \sum_{\boldsymbol j,\boldsymbol j'}\left( \mathcal E({\boldsymbol r}_{\boldsymbol j}) \right)^*\left(\mathcal H-\Delta_d\mathbb 1\right)^{-1}_{\boldsymbol j,\boldsymbol j'} \delta_{j'_z,1}(v_d)_{\boldsymbol j_\perp'} \right|^2
\\ =&2 |\tilde c(t)|^2 \Omega^2 \Gamma \left| \frac{\gamma_d}{e^{2ik_0L}\Gamma^2-(\Gamma+\gamma_d)^2} \right|^2,
\end{aligned}
\end{equation*}
where in the second line we replace the Green's function by its paraxial counterpart as in Eq.~\eqref{eq:decompG} and used Eq.~\eqref{eq:pargreen}, and in the third line restricted $(\mathcal H-\Delta_d\mathbb 1)$ to the subspace of symmetric and anti-symmetric Gaussian states, replaced $\sum_{\boldsymbol j_\perp}\approx\int d{\boldsymbol r_\perp}/\delta_\perp^2$ and used the property of Eq.~\eqref{eq:phaseprofile}. We note that $P_\rightarrow$ vanishes for $\gamma_d\approx0$. Similarly, the flux of photons emitted in the Gaussian mode $\mathcal E^*({\boldsymbol r})$  (i.e., propagating to the left), reads
\begin{equation*}
\begin{aligned}
P_\leftarrow(t)=&c \left| \int d{\boldsymbol r}_\perp  \mathcal E({\bf r}) \psi({\boldsymbol r},t) \right|^2
\\ =&2 |\tilde c(t)|^2 \Omega^2 \Gamma \left| \frac{e^{2ik_0L} \Gamma-(\Gamma+\gamma_d)}{e^{2ik_0L}\Gamma^2-(\Gamma+\gamma_d)^2} \right|^2.
\end{aligned}
\end{equation*}
For $\gamma_d\approx0$ this reduces to $P_\leftarrow(t)=\tilde \gamma |\tilde c(t)|^2$, showing that the photon is emitted in the left-propagating Gaussian mode. This allows to perform a transfer from the quantum memory state $\ket{\psi_i}$ to a propagating photonic qubit, as well as the time-reversed process.
  
\subsection{Beyond the Markov approximation}
Finally, we discuss the effects of time-delays in the atomic dynamics, arising from the finite propagation time of photons between arrays, which were neglected in the previous sections. These effects become relevant only when the arrays are separated by $L\gtrsim c/\Gamma \sim10$m for $\Gamma$ in the MHz range, which can be realized by mediating photons exchanged between arrays with optical lenses or fibers. We first show how this affects the decay rates of dark and bright states, and then study its effect on the state transfer fidelity.

\subsubsection{Effect on dark and bright states}
Let us write the state of the system
\begin{equation*}
\ket{\psi(t)}=\sum_{{\boldsymbol j}}c_{\boldsymbol j}(t)\sigma_{\boldsymbol j}^+\ket{\mathcal G}\ket{0}+\int d{\boldsymbol k}\sum_\lambda \psi_\lambda({\boldsymbol k},t)\ket{\mathcal G}\ket{{\boldsymbol k},\lambda},
\end{equation*}  
and integrate the field dynamics without neglecting retardation in photon propagation between different arrays, yielding \cite{Grankin2018}, with the definitions of Eqs.~\eqref{eq:defH0} and~\eqref{eq:defH1}
\begin{equation*}\begin{aligned}
\dot c_{(\boldsymbol j_\perp,1)}(t)=-i \sum_{\boldsymbol j'_\perp}&\Big[(\mathcal H_{0})_{\boldsymbol j_\perp,\boldsymbol j'_\perp}c_{(\boldsymbol j'_\perp,1)}(t)
\\ &+e^{-\kappa L}(\mathcal H_{1})_{\boldsymbol j_\perp,\boldsymbol j'_\perp}c_{(\boldsymbol j'_\perp,2)}(t-\tau)\Big],
\\\dot c_{(\boldsymbol j_\perp,2)}(t)=-i \sum_{\boldsymbol j'_\perp}&\Big[(\mathcal H_{0})_{\boldsymbol j_\perp,\boldsymbol j'_\perp}c_{(\boldsymbol j'_\perp,2)}(t)
\\ &+e^{-\kappa L}(\mathcal H_{1})_{\boldsymbol j_\perp,\boldsymbol j'_\perp}c_{(\boldsymbol j'_\perp,1)}(t-\tau)\Big],
\end{aligned}\end{equation*}
with $\tau=L/c$. Here we added an attenuation coefficient $\kappa$ accounting for additional decay channels induced by possible optical elements mediating the exchanged photons, but neglected any effect of dispersion. In particular, for the dark and bright state amplitudes we get
\begin{equation*}
\dot c_{b/d}(t)= -\frac{\Gamma+\gamma_d}{2}c_{b/d}(t)\mp\frac{\Gamma}2 e^{-\kappa L}c_{b/d}(t-\tau).
\end{equation*}

Defining the Laplace transform variables $\tilde{c}(s)=\mathcal{L}[c(t)](s)$,
we arrive at 
\begin{equation*}
\label{eq:laplacesys}\tilde{c}_{b/d}(s)= \Big(s+\frac{\Gamma(1-e^{-s\tau-\kappa L})+\gamma_d}{2}\Big)^{-1}c_{b/d}(0),
\end{equation*}
which cannot be analytically inverted directly. An analytical approximation
can however be obtained by expanding $e^{-s\tau}=1-s\tau+\mathcal{O}(s\tau)^{2}$
to lowest order in $\Gamma\tau$, yielding 
\begin{equation}
\begin{aligned}c_{b}(t)= & 2\frac{\text{exp}\left(-\frac{(1+e^{-\kappa L})\Gamma+\gamma_d}{2-\Gamma\tau e^{-\kappa L}}t\right)}{2-\Gamma\tau e^{-\kappa L}}c_{b}(0)\\
c_{d}(t)= &2 \frac{\text{exp}\left(-\frac{(1-e^{-\kappa L})\Gamma+\gamma_d}{2+\Gamma\tau e^{-\kappa L}}t\right)}{2+\Gamma\tau e^{-\kappa L}}c_{d}(0).
\end{aligned}
\label{eq:firstorderapprox}
\end{equation}
From Eq.~\eqref{eq:firstorderapprox} increasing 
the retardation $\Gamma\tau$ {decreases} the decay of the dark
state due to atomic losses with rate $\gamma_d$. This is a consequence
of the dark state being now a superposition of field and atomic excitations,
with only the atomic part decaying if $\kappa L=0$. If we include a finite attenuation $\kappa L$, the photonic component also induces losses.

The photonic field is expressed as
\begin{equation*}\label{eq:psirtnonmarkov}
\begin{aligned}
\psi({\boldsymbol r},t)=&-i\sqrt{\frac{\gamma_ek_0^2}{6\pi c}} \sum_{\boldsymbol j}c_{\boldsymbol j}(t-\tau|{\boldsymbol r-\boldsymbol r_{\boldsymbol j}}|/L){G}({\boldsymbol r}-{\boldsymbol r}_{\boldsymbol j}),
\end{aligned}
\end{equation*}
providing for the photonic flux $E(z,t)=c\int d{\boldsymbol r_\perp} |\psi({\boldsymbol r},t)|^2$, within the paraxial approximation for the Green's tensor,
\begin{equation*}\label{eq:psirtnonmarkov}
\begin{aligned}
E&(z,t)\approx\frac{\Gamma}{4} \int d{\boldsymbol r_\perp}
\\&\Big| \mathcal E({\boldsymbol r})\left[c_b(t-\tau|{z-z_1}|/L)+c_d(t-\tau|{z-z_1}|/L)\right]
\\ &+ \mathcal E^*({\boldsymbol r})\left[c_b(t-\tau|{z-z_2}|/L)-c_d(t-\tau|{z-z_2}|/L)\right]\Big|^2,
\end{aligned}
\end{equation*}
where $z_1=-L/2$ and $z_2=L/2$ denote the position of the first and second arrays along $z$. In particular, in the dark state the system can reach a quasi-equilibrium if the exponent in the decay of the dark state amplitude in Eq.~\eqref{eq:firstorderapprox} is much smaller than $1/\tau$, and we get (with $z_R\gg L$ for simplicity)
$E(z,t)\approx{\Gamma} \sin(k_0z)^2|c_d(t)|^2$. While this quantity shows that the arrays continuously exchange photons at a rate $\sim \Gamma$ even when $\Gamma\tau\to0$, the total number of photons between the arrays at any time, given by
\begin{equation*}
N_\text{ph}=\frac{1}{c}\int_{z_1}^{z_2}E(z,t) dz = \frac{\Gamma \tau}2 |c_d(t)|^2,
\end{equation*}
vanishes in that limit.

\subsubsection{Effect on state transfer fidelity}
The effect of retardation on the state transfer fidelity can be studied in a similar way. Following the same procedure, with the notations of Sec.~\ref{sec:solutioneffectivemodel}, we have
\begin{equation*}
\begin{aligned}
\dot c_{1}\left(t\right)  =&-i\frac{\Omega}{\sqrt{2}}\left( c_{b}\left(t\right)+c_{d}\left(t\right)\right)\\
\dot c_{2}\left(t\right)  =&-i\frac{\Omega}{\sqrt{2}}\left( c_{b}\left(t\right)-c_{d}\left(t\right)\right) \\
\dot c_{b}\left(t\right)  =& -\frac{\Gamma+\gamma_d}{2}c_{b}(t)-\frac{\Gamma}2 e^{-\kappa L}c_{b}(t-\tau)
\\&-i\frac{\Omega}{\sqrt{2}}\left( c_{1}\left(t\right)+c_{2}\left(t\right)\right) \\
\dot c_{d}\left(t\right)  =& -\frac{\Gamma+\gamma_d}{2}c_{d}(t)+\frac{\Gamma}2 e^{-\kappa L}c_{d}(t-\tau)
\\&-i\frac{\Omega}{\sqrt{2}}\left( c_{1}\left(t\right)-c_{2}\left(t\right)\right),
\end{aligned}
\end{equation*}
with initial conditions $c_1(0)=1$, $c_2(0)=c_d(0)=c_b(0)=0$.
The solution for the Laplace transform of $c_2$ then reads
\begin{equation*}
\begin{aligned}\tilde{c}_{2}(s)= & \frac{2\Omega^{2}\Gamma e^{-s\tau-\kappa L}}{\left(2\Omega^{2}+s(2s+\Gamma+\gamma_d)\right)^{2}-s^{2}\Gamma^{2}e^{-2s\tau-2\kappa L}},
\end{aligned}
\end{equation*}
which is Laplace-inverted numerically.

\begin{figure}
\includegraphics[width=0.5\textwidth]{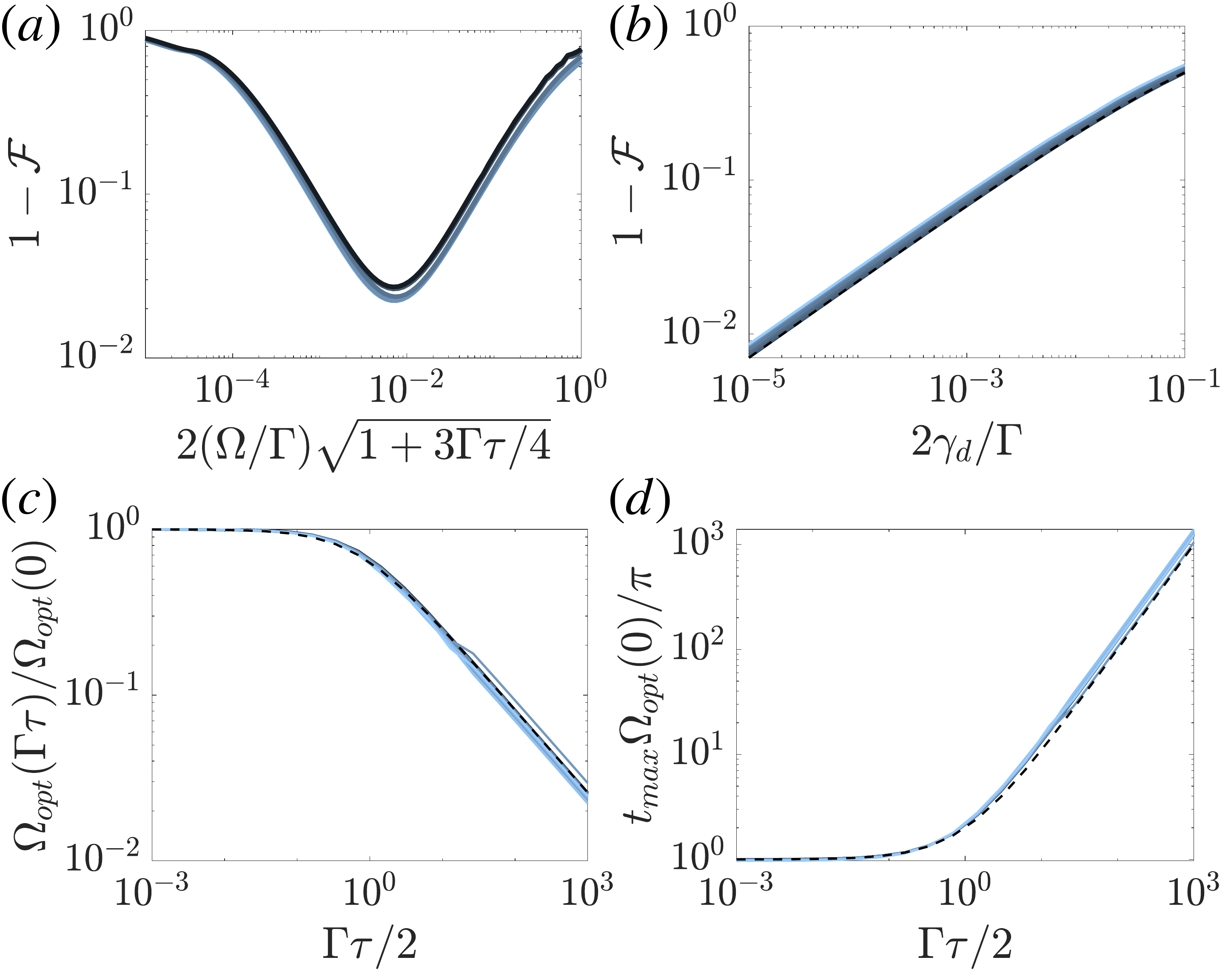}
\caption{\label{fig:delay}(a)~Infidelity for state transfer, with $2\gamma_d=10^{-4} \Gamma$, $\Gamma\tau/2\in[10^{-3},10^{3}]$
(light to dark blue) and $\kappa=0$.
(b)~Infidelity for state transfer, with optimal $\Omega$, $(\Gamma\tau/2)\in[10^{-3},10^{3}]$
(light to dark blue) and $\kappa=0$. Dashed black: $1-\text{exp}(-\pi\sqrt{2\gamma_{d}/\gamma_{b}})$.
(c)~Optimal value of $\Omega$, with $2\gamma_d/\Gamma\in[10^{-5},10^{-1}]$
(light to dark blue) and $\kappa=0$. Dashed black: $1/\sqrt{1+3\Gamma\tau/4}$. (d)~Transfer time $t_\text{max}$, with optimal $\Omega$, $2\gamma_d/\Gamma\in[10^{-5},10^{-1}]$
(light to dark blue) and $\kappa=0$. Dashed black: $1+\Gamma\tau/2$.}
\end{figure}

The effect of retardation is represented for $\kappa=0$ in Fig.~\ref{fig:delay}
with $\Gamma\tau\in[10^{-3},10^{3}]$ and $\gamma_d/\Gamma\in[10^{-5},10^{-1}]$.
In all these figures we see that the effect is to rescale
the parameters of the system. In Figs.~\ref{fig:delay}(a,b)
we see in particular that the fidelity for quantum state transfer
is almost constant for all values of the retardation $\Gamma\tau$. This can be understood as while increasing $\Gamma\tau$ increases the state transfer time $t_\text{max}$, the decay rate of the dark state in Eq.~\eqref{eq:firstorderapprox} decreases, resulting in a constant overall loss probability.
In Fig.~\ref{fig:delay}(c,d) we see that the required
optimal $\Omega$ decreases, such that the transfer time $t_\text{max}$ increases
linearly with the delay at large $\Gamma\tau$. For finite attenuation $\kappa L$ this reduces the transfer fidelity.

\section{Effect of finite Lamb-Dicke parameter and temperature}

Here we discuss the effects of phononic degrees of freedom for atoms trapped in optical lattices with finite Lamb-Dicke parameter $\eta$, and thermal phonon distribution with mean number $n_\text{th}$. We derive a correction to $\mathcal H$, and in particular show that the
spread of the atomic wavefunction leads to an additional individual
decay of each atom of $\gamma_{e}\eta^{2}\left(2n_\text{th}+1\right)$.

\subsection{Model}
Assuming that each atom
is trapped with a harmonic potential with frequency $\omega_{v}$,
the non-hermitian Hamiltonian $H_\text{dip}$, including now the coupling to motional degrees of freedom, reads
\begin{equation}
\begin{aligned}
 H_\text{dip} =&\omega_{v}\sum_{\boldsymbol{j},\alpha}a_{\boldsymbol{j},\alpha}^{\dagger}a_{\boldsymbol{j},\alpha}
 \\ &-\frac{i\gamma_{e}}{2}\sum_{\boldsymbol{j},\boldsymbol{j}^{\prime}}\sigma_{\boldsymbol{j}}^{+}\sigma_{\boldsymbol{j}^{\prime}}^{-}{G}\left({\boldsymbol r}_{\boldsymbol{j}}-{\boldsymbol r}_{\boldsymbol j'}+\hat{\boldsymbol{r}}_{\boldsymbol{j}}-\hat{\boldsymbol{r}}_{\boldsymbol{j}^{\prime}}\right)  
 \\ =&\omega_{v}\sum_{\boldsymbol{j},\alpha}a_{\boldsymbol{j},\alpha}^{\dagger}a_{\boldsymbol{j},\alpha}
  -\frac{i\gamma_{e}}{2}\sum_{\boldsymbol{j},\boldsymbol{j}^{\prime}}\sigma_{\boldsymbol{j}}^{+}\sigma_{\boldsymbol{j}^{\prime}}^{-}{G}\left({\boldsymbol r}_{\boldsymbol{j}}-{\boldsymbol r}_{\boldsymbol j'}\right)  
  \\ &-\frac{i\gamma_{e}}{2}\sum_{\boldsymbol{j},\boldsymbol{j}^{\prime}}\sigma_{\boldsymbol{j}}^{+}\sigma_{\boldsymbol{j}^{\prime}}^{-}\left[\left(\hat{\boldsymbol{r}}_{\boldsymbol{j}}-\hat{\boldsymbol{r}}_{\boldsymbol{j}^{\prime}}\right)\cdot\boldsymbol{\nabla}\right]{G}\left({\boldsymbol r}_{\boldsymbol{j}}-{\boldsymbol r}_{\boldsymbol j'}\right)  
   \\ &-\frac{i\gamma_{e}}{4}\sum_{\boldsymbol{j},\boldsymbol{j}^{\prime}}\sigma_{\boldsymbol{j}}^{+}\sigma_{\boldsymbol{j}^{\prime}}^{-}\left[\left(\hat{\boldsymbol{r}}_{\boldsymbol{j}}-\hat{\boldsymbol{r}}_{\boldsymbol{j}^{\prime}}\right)\cdot\boldsymbol{\nabla}\right]^2{G}\left({\boldsymbol r}_{\boldsymbol{j}}-{\boldsymbol r}_{\boldsymbol j'}\right)  
   \\&+ \mathcal O(\eta\sqrt{2n_\text{th}+1})^3,
   \label{eq:H_nh_motion}
\end{aligned}
\end{equation}
where we performed a Taylor expansion for the Green's tensor, with $\eta=k_0/\sqrt{2m\omega_v}$ the Lamb-Dicke parameter, $a_{\boldsymbol{j},\alpha}$ the annihilation operator of
the motional excitation of atom ${\boldsymbol j}$ along axis $\alpha\in(x,y,z)$, $\hat{\boldsymbol {r}}_{\boldsymbol{j}}$
the quantized coordinates of atom $\boldsymbol{j}$ relative
to its trap center position ${\boldsymbol {r}}_{\boldsymbol{j}}$, and where ${\boldsymbol \nabla}$ acts on $ {G}({\boldsymbol r})$.  We assumed the unperturbed density matrix of the system factorizes as $\rho^{\left(0\right)}=\rho_\text{at}\otimes\rho_\text{th}$ where
$\rho_\text{th}$ stands for a thermal distribution of phononic modes with mean number $n_\text{th}$, while $\rho_\text{at}$ accounts for the internal atomic degrees of freedom.

\subsection{Elimination of phonon modes}
We now perform an adiabatic elimination of the motional degrees
of freedom in Eq.~\eqref{eq:H_nh_motion}. We write $H_\text{dip}=H_0+V$ as a sum of a free Hamiltonian $H_0$ and an interaction term $V$. Assuming $\eta\sqrt{2n_\text{th}+1}\ll 1$, we can truncate the expansion in Eq. (\ref{eq:H_nh_motion})
to second order. We thus have
\begin{equation*}
\begin{aligned}
H_{0}= &\omega_{v}\sum_{\boldsymbol{j},\alpha}a_{\boldsymbol{j},\alpha}^{\dagger}a_{\boldsymbol{j},\alpha}
 -\frac{i\gamma_{e}}{2}\sum_{\boldsymbol{j},\boldsymbol{j}^{\prime}}\sigma_{\boldsymbol{j}}^{+}\sigma_{\boldsymbol{j}^{\prime}}^{-}{G}\left({\boldsymbol r}_{\boldsymbol{j}}-{\boldsymbol r}_{\boldsymbol j'}\right)  
\\V =&  -\frac{i\gamma_{e}}{2}\sum_{\boldsymbol{j},\boldsymbol{j}^{\prime}}\sigma_{\boldsymbol{j}}^{+}\sigma_{\boldsymbol{j}^{\prime}}^{-}\left[\left(\hat{\boldsymbol{r}}_{\boldsymbol{j}}-\hat{\boldsymbol{r}}_{\boldsymbol{j}^{\prime}}\right)\cdot\boldsymbol{\nabla}\right]{G}\left({\boldsymbol r}_{\boldsymbol{j}}-{\boldsymbol r}_{\boldsymbol j'}\right)  
   \\ &-\frac{i\gamma_{e}}{4}\sum_{\boldsymbol{j},\boldsymbol{j}^{\prime}}\sigma_{\boldsymbol{j}}^{+}\sigma_{\boldsymbol{j}^{\prime}}^{-}\left[\left(\hat{\boldsymbol{r}}_{\boldsymbol{j}}-\hat{\boldsymbol{r}}_{\boldsymbol{j}^{\prime}}\right)\cdot\boldsymbol{\nabla}\right]^2{G}\left({\boldsymbol r}_{\boldsymbol{j}}-{\boldsymbol r}_{\boldsymbol j'}\right).  
\end{aligned}
\end{equation*}
Moving to an interaction picture with respect to $H_{0}$, we get an effective Hamiltonian $\tilde H$ to second order perturbation in $V$, assuming further that $\gamma\eta\sqrt{2n_\text{th}+1}\ll\omega_v$. We then obtain
\begin{equation}
\tilde H=\text{Tr}_\text{ph}\left[V\left(t\right)\rho_\text{th}\right]-i\int_{-\infty}^{t}ds\text{Tr}_\text{ph}\left[V\left(t\right)V\left(s\right)\rho_\text{th}\right],\label{eq:H_eff}
\end{equation}
where we denote $\text{Tr}_\text{ph}$ for the trace over phononic degrees of freedom. 

The first term in Eq.~\eqref{eq:H_eff} is evaluated as
\begin{equation*}
\begin{aligned}
 & \text{Tr}_\text{ph}\left[V\left(t\right)\rho_\text{th}\right]\\
 & =-\frac{i\gamma_{e}}{4}\text{Tr}_\text{ph}\left[\sum_{\boldsymbol{j},\boldsymbol{j}^{\prime}}\sigma_{\boldsymbol{j}}^{+}\sigma_{\boldsymbol{j}^{\prime}}^{-}\left[\left(\hat{\boldsymbol{r}}_{\boldsymbol{j}}-\hat{\boldsymbol{r}}_{\boldsymbol{j}^{\prime}}\right)\cdot\boldsymbol{\nabla}\right]^2{G}\left({\boldsymbol r}_{\boldsymbol{j}}-{\boldsymbol r}_{\boldsymbol j'}\right)  
\rho_\text{th}\right]\\
 & =-\frac{i\gamma_{e}}{2k_0^2}\eta^{2}\left(2n_\text{th}+1\right)\sum_{\boldsymbol{j}\neq\boldsymbol{j}^{\prime}} \sigma_{\boldsymbol{j}}^{+}\sigma_{\boldsymbol{j}^{\prime}}^{-} ({\boldsymbol \nabla}\cdot \boldsymbol \nabla){G}\left({\boldsymbol r}_{\boldsymbol{j}}-{\boldsymbol r}_{\boldsymbol j'}\right).
\end{aligned}
\end{equation*}
Using the vector field identity ${\boldsymbol \nabla}\times(\boldsymbol{\nabla}\times\boldsymbol{V})=\boldsymbol{\nabla}\left(\boldsymbol{\nabla}\cdot\boldsymbol V\right)-(\boldsymbol{\nabla}\cdot\boldsymbol{\nabla})\boldsymbol V$ and Eq.~\eqref{eq:maxwell}, we further get
\begin{equation*}
\begin{aligned}
& \text{Tr}_\text{ph}\left[V\left(t\right)\rho_\text{th}\right] \\
  &=\frac{i\gamma_{e}\eta^{2}}{2}\left(2n_\text{th}+1\right)\sum_{\boldsymbol{j}\neq\boldsymbol{j}^{\prime}}\sigma_{\boldsymbol{j}}^{+}\sigma_{\boldsymbol{j}^{\prime}}^{-} {G}\left({\boldsymbol r}_{\boldsymbol{j}}-{\boldsymbol r}_{\boldsymbol j'}\right) \nonumber\\
 &\! -\!\frac{i\gamma_{e}}{2k_0^2}\eta^{2}\!\left(2n_\text{th}+1\right)\!\sum_{\boldsymbol{j}\neq\boldsymbol{j}^{\prime}}\sigma_{\boldsymbol{j}}^{+}\sigma_{\boldsymbol{j}^{\prime}}^{-}\boldsymbol{p}^{*}\cdot\! \left({\boldsymbol \nabla}\!\left({\boldsymbol \nabla}\!\cdot\!\hat{\boldsymbol G}\left({\boldsymbol r}_{\boldsymbol{j}}-{\boldsymbol r}_{\boldsymbol{j}'}\right)\right)\right)\cdot\! \boldsymbol{p}.
\end{aligned}
\end{equation*}
The last term in this equation stands for the variation of the longitudinal
part of the Green's tensor, and introduces a renormalization of the
near-field interaction. Dropping this near-field term, which decays as $|\boldsymbol r_{\boldsymbol j}-\boldsymbol r_{\boldsymbol j'}|^3$, we finally have
\begin{equation}\label{eq:TrVresult}
\begin{aligned}
 \text{Tr}_\text{ph}\left[V\left(t\right)\rho_\text{th}\right] = & -\frac{i\gamma_{e}}{2}\eta^{2}\left(2n_\text{th}+1\right)\sum_{\boldsymbol{j}}\sigma_{\boldsymbol{j}}^{+}\sigma_{\boldsymbol{j}}^{-}
\\+&\frac{i\gamma_{e}}{2}\eta^{2}\left(2n_\text{th}+1\right)\sum_{\boldsymbol{j},\boldsymbol{j}^{\prime}}\sigma_{\boldsymbol{j}}^{+}\sigma_{\boldsymbol{j}^{\prime}}^{-}{G}\left({\boldsymbol r}_{\boldsymbol{j}}-{\boldsymbol r}_{\boldsymbol j'}\right) 
\end{aligned}
\end{equation}

Using the identity $\int_{-\infty}^{t}ds\text{Tr}_\text{ph}\left[\hat{\alpha}_{\boldsymbol{j}}\left(t\right)\hat{\alpha}_{\boldsymbol{j}^{\prime}}^{\prime}\left(s\right)\rho_\text{th}\right]=\delta_{\alpha,\alpha^{\prime}}\delta_{\boldsymbol{j},\boldsymbol{j}^{\prime}}/\left(i\omega_{v}\right)$, with $\hat \alpha_{\boldsymbol j}$ the component of $\hat {\boldsymbol r}_{\boldsymbol j}$ along axis $\alpha$,
we get for the second term in Eq.~\eqref{eq:H_eff}
\begin{equation}
\begin{aligned}
  -i&\int_{-\infty}^{t}ds\text{Tr}_\text{ph}\left[V\left(t\right)V\left(s\right)\rho_\text{th}\right] \\
=  -&\frac{\gamma_{e}^{2}\eta^{2}}{4\omega_{v}k_0^2}\!\sum_{\boldsymbol{j},\boldsymbol{j}^{\prime},{\boldsymbol j''}\neq\{\boldsymbol j,\boldsymbol j'\}}\!\!\!\!\!\!\!\!\!\!{\boldsymbol \nabla}G\left({\boldsymbol r}_{\boldsymbol{j}}-{\boldsymbol r}_{\boldsymbol{j}''}\right)\cdot \!\!{\boldsymbol \nabla}G\left({\boldsymbol r}_{\boldsymbol{j}''}\!-{\boldsymbol r}_{\boldsymbol{j}'}\right)\!\sigma_{\boldsymbol{j}}^{+}\!\sigma_{\boldsymbol{j}'}^{-}.\label{eq:VtVs}
\end{aligned}
\end{equation}
This term can be evaluated in the limit of infinite arrays ($N_\perp\to\infty$), by splitting the sum as 
\begin{equation*}
\sum_{\boldsymbol j,\boldsymbol j',\boldsymbol j''\neq \{\boldsymbol j,\boldsymbol j'\}}=\sum_{\boldsymbol j,\boldsymbol j',\boldsymbol j''}(1-\delta_{\boldsymbol j,\boldsymbol j''}-\delta_{\boldsymbol j',\boldsymbol j''}+\delta_{\boldsymbol j,\boldsymbol j''}\delta_{\boldsymbol j',\boldsymbol j''}).
\end{equation*}
Using the representation of Eq.~\eqref{eq:Gintdq} for the Green's tensor and the property of Eq.~\eqref{eq:sumgdef}, we then find that 
the dark and bright states $(1/\sqrt{2N})\sum_{\boldsymbol j_\perp}(\sigma^+_{(\boldsymbol j_\perp,1)}\pm\sigma^+_{(\boldsymbol j_\perp,2)})\ket{\mathcal G}$ are eigenstates of the Hamiltonian in Eq.~\eqref{eq:VtVs} with purely real eigenvalues, provided $\delta_\perp<\lambda_0$ and $k_0L=m\pi$. Therefore, for these states the term in Eq.~\eqref{eq:VtVs} contributes only a frequency renormalization.

Thus, only the term of Eq.~\eqref{eq:TrVresult} contributes to the radiation, such that the decay rates are corrected as
 \begin{equation*}
\begin{aligned}
\gamma_{d/b}\to \gamma_e\eta^{2}\left(2n_\text{th}+1\right)+\gamma_{d/b}\left(1-\eta^{2}\left(2n_\text{th}+1\right)\right)
\end{aligned}
\end{equation*}
This shows an additional individual decay for each
atom, and an additional rescaling of the interatomic interaction. In order have a given ratio for $\gamma_d/\gamma_b$, we must thus satisfy the condition $\eta^2(2n_\text{th}+1)\lesssim \gamma_d/\gamma_b$.

\section{Effect of missing atoms}
Here we discuss how the presence of defects in the atomic arrays affect the system.
The effect of holes can be accounted for by writing for the dipole-dipole interaction Hamiltonian of Eq.~\eqref{mathcalH}
\begin{equation}\label{eq:mathcalHholes}
\mathcal H=\mathcal H^\text{ideal}-\mathcal H^\text{holes},
\end{equation}
where $\mathcal H^\text{ideal}$ is the matrix without defects, and $\mathcal H^\text{holes}_{\boldsymbol j,\boldsymbol j'}=\mathcal H^\text{ideal}_{\boldsymbol j,\boldsymbol j'}$ if atom $\boldsymbol j$ or $\boldsymbol j'$ is missing and $\mathcal H^\text{holes}_{\boldsymbol j,\boldsymbol j'}=0$ otherwise.

\begin{figure}
\includegraphics[width=0.4\textwidth]{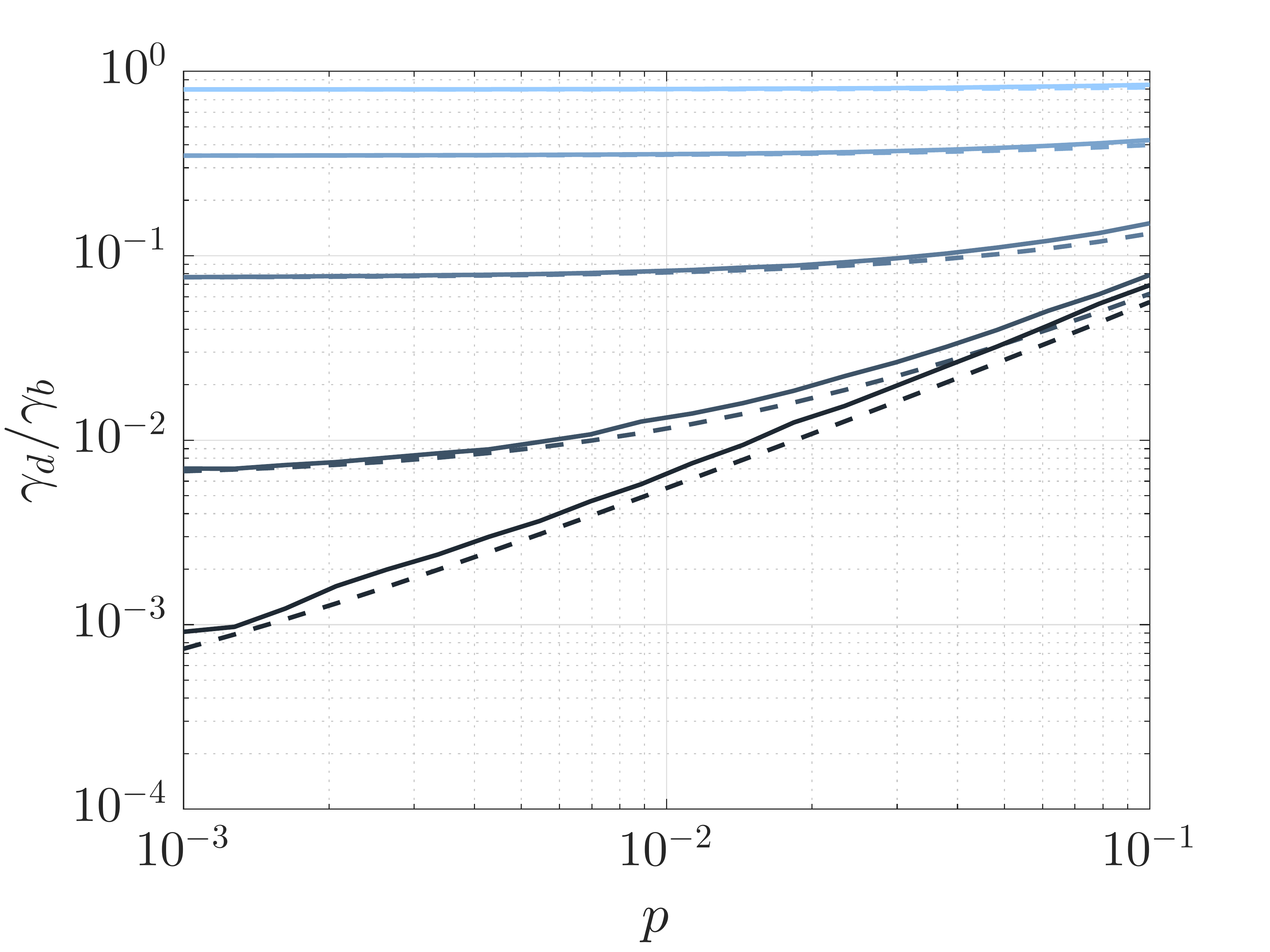}
\caption{\label{fig:defects} Effect of probability $p$ of having defects on each site, with $N_\perp=4,8,12,16,20$ (light to dark blue), $\delta_\perp=0.5$, $L=30\lambda_0$. Solid: numerics. Dashed: expression from Eq.~\eqref{eq:defects}. }
\end{figure}

We first consider the situation where a single atom, say atom $\boldsymbol i$, is missing from the arrays. The dipole-dipole interaction Hamiltonian can be expressed from Eq.~\eqref{eq:mathcalHholes} as
\begin{equation*}
\begin{aligned}
&H_\text{dip}=-i\frac{\gamma_e}2\sum_{\boldsymbol j,\boldsymbol j'}\sigma_{\boldsymbol j}^+\sigma_{\boldsymbol j'}^- G({\boldsymbol r}_{\boldsymbol j}-{\boldsymbol r}_{\boldsymbol j'})
-i\frac{\gamma_e}2\sigma_{\boldsymbol i}^+\sigma_{\boldsymbol i}^- 
\\&+i\frac{\gamma_e}2\sum_{\boldsymbol j'}\sigma_{\boldsymbol i}^+\sigma_{\boldsymbol j'}^- G({\boldsymbol r}_{\boldsymbol i}-{\boldsymbol r}_{\boldsymbol j'})
+i\frac{\gamma_e}2\sum_{\boldsymbol j}\sigma_{\boldsymbol j}^+\sigma_{\boldsymbol i}^- G({\boldsymbol r}_{\boldsymbol j}-{\boldsymbol r}_{\boldsymbol i}).
\end{aligned}\end{equation*} 
For the state \mbox{$\ket{\psi_n}\propto\sum_{\boldsymbol j\neq\boldsymbol i}(c_n)_{\boldsymbol j}\sigma_{\boldsymbol j}^+\ket{\mathcal G}$}, we then obtain
\begin{equation*}\label{eq:guess1hole}
\begin{aligned}
\bra{\psi_n}H_\text{dip}\ket{\psi_n}=&\left(\Delta_n-i\frac{\gamma_n}2\right)\left(1-|(c_n)_{\boldsymbol i}|^2\right)
\\&-i\frac{\gamma_e}2\left|(c_n)_{\boldsymbol i}\right|^2+\mathcal O(\left|(c_n)_{\boldsymbol i}\right|^4).
\end{aligned}
\end{equation*}

Let us now consider a situation where each atom ${\boldsymbol i}$ has a probability $0\leq p \leq 1$ of being missing, and denote the 
associated random variable as $s_{\boldsymbol i}$. We then have
\begin{equation*}
\begin{aligned}
&H_\text{dip}=-i\frac{\gamma_e}2\sum_{\boldsymbol j,\boldsymbol j'}\sigma_{\boldsymbol j}^+\sigma_{\boldsymbol j'}^- G({\boldsymbol r}_{\boldsymbol j}-{\boldsymbol r}_{\boldsymbol j'})
-i\frac{\gamma_e}2\sum_{\boldsymbol i}s_{\boldsymbol i}\sigma_{\boldsymbol i}^+\sigma_{\boldsymbol i}^- 
\\&+i\frac{\gamma_e}2\sum_{\boldsymbol j'}s_{\boldsymbol i}\sigma_{\boldsymbol i}^+\sigma_{\boldsymbol j'}^- G({\boldsymbol r}_{\boldsymbol i}-{\boldsymbol r}_{\boldsymbol j'})
+i\frac{\gamma_e}2\sum_{\boldsymbol j}s_{\boldsymbol i}\sigma_{\boldsymbol j}^+\sigma_{\boldsymbol i}^- G({\boldsymbol r}_{\boldsymbol j}-{\boldsymbol r}_{\boldsymbol i})
\\ &-i\frac{\gamma_e}2\sum_{\boldsymbol i\neq\boldsymbol i'}s_{\boldsymbol i}s_{\boldsymbol i'}\sigma_{\boldsymbol i}^+\sigma_{\boldsymbol i'}^- G({\boldsymbol r}_{\boldsymbol i}-{\boldsymbol r}_{\boldsymbol i'}) .
\end{aligned}\end{equation*} 
Denoting $\overline{\phantom{aa}}$ for the statistical average, and using here $\overline{s_{\boldsymbol i}}=p$ and $\overline{s_{\boldsymbol i}s_{\boldsymbol i'}}=p^2+\delta_{\boldsymbol i,\boldsymbol i'}(p-p^2)$, we obtain for  \mbox{$\ket{\psi_n}\propto\sum_{\boldsymbol j}(c_n)_{\boldsymbol j}(1-s_{\boldsymbol j})\sigma_{\boldsymbol j}^+\ket{\mathcal G}$}
\begin{equation}
\begin{aligned}\label{eq:defects}
\overline{\bra{\psi_n}H_\text{dip}\ket{\psi_n}}=&\left(\Delta_n-i\frac{\gamma_n}2\right)\left(1-p\right)-i\frac{\gamma_e}2p
\\&+ \mathcal O(p^2).
\end{aligned}
\end{equation}
In Fig.~\ref{fig:defects} we show the agreement between this expression and numerical simulations, where the dark and bright states decay rates are averaged over $100$ realizations of $s_{\boldsymbol i}$. In order to achieve a given ratio for $\gamma_d/\gamma_b$, we must thus have $p\lesssim \gamma_d/\gamma_b$.

\section{Multiple excitations}
For states with more than one atom in $\ket{e}$ or $\ket{s}$ the fact that each atom cannot support more than a single excitation generates an atomic non-linearity, which induces an additional decay rate. The state $\ket{\psi_d^{(2)}}\propto(\sigma_d^+)^2\ket{\mathcal G}$ for instance is not an eigenstate of $H_\text{dip}$, however we can treat this non-linearity in first order perturbation theory. We get 
\begin{equation}
\begin{aligned}
\bra{\psi^{(2)}_d}H_\text{dip}\ket{\psi_d^{(2)}}&=\left(\Delta_d-i\frac{\gamma_d}{2}\right)\left(1-\sum_{\boldsymbol j_\perp}|(v_d)_{\boldsymbol j}|^4\right)
\\ &+\left(\Delta_d-i\frac{\gamma_d}{2}\right)-i\frac{\gamma_e}2\sum_{\boldsymbol j_\perp}|(v_d)_{\boldsymbol j_\perp}|^4.
\end{aligned}\label{eq:Hdip2exc}
\end{equation}
The decay rate per excitation, as represented in Fig.~\ref{fig:fig2}(a), is then obtained as 
\begin{equation*}
\gamma^{(2)}=-\text{Im}\left(\bra{\psi^{(2)}_d}H_\text{dip}\ket{\psi_d^{(2)}}\right).
\end{equation*}
From Eq.~\eqref{eq:defects}, we can interpret the result of Eq.~\eqref{eq:Hdip2exc} as one of the two excitations decays with rate $\gamma_d$, and acts as a defect for the other excitation with probability $p=\sum_{\boldsymbol j_\perp}|(v_d)_{\boldsymbol j_\perp}|^4$ identified as twice the inverse participation ratio of the dark state.

\section{Implementation with four-level atoms}
\begin{figure}
\includegraphics[width=0.5\textwidth]{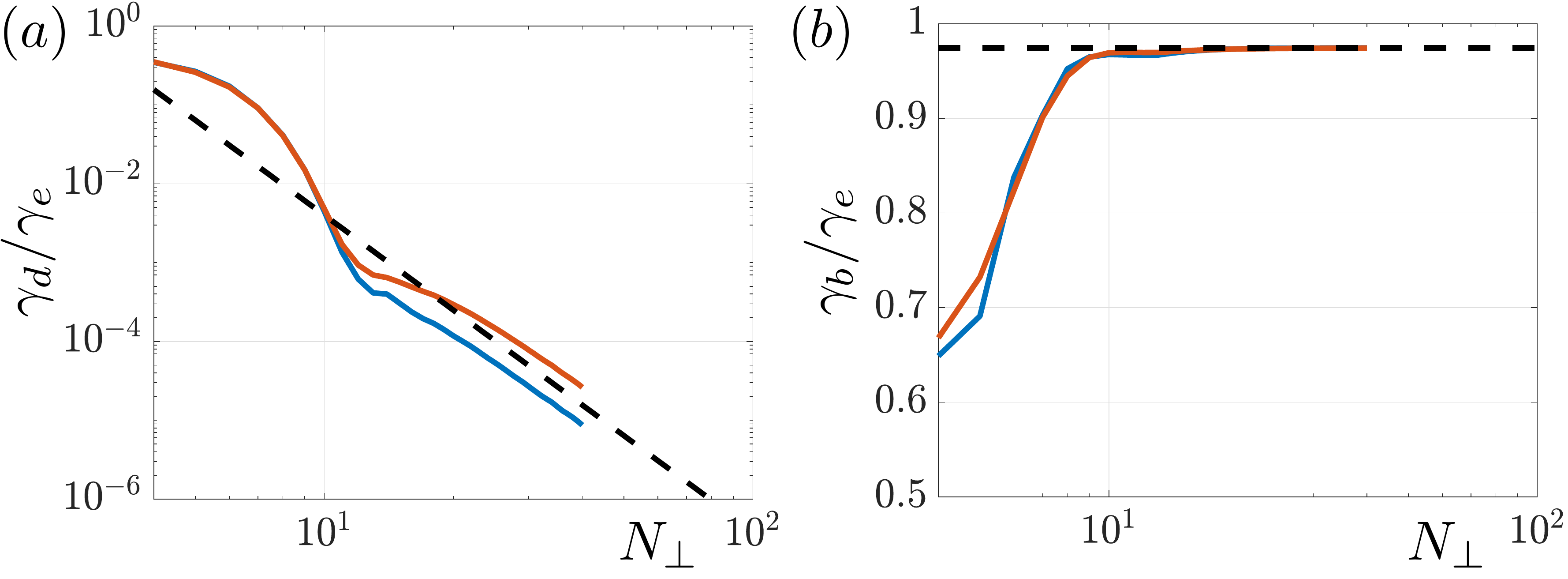}
\caption{\label{fig:FigIso}(a)~Dark and (b)~bright state decay rates for two-level atoms with circular transition (blue) and four-level atoms (red), with $L=20\lambda_0$, $\delta_\perp=0.7\lambda_0$. Dashed black: (a)~$\propto1/N_\perp^4$ and (b)~$2\Gamma/\gamma_e$.}
\end{figure}

In all the calculations above and in the main text we treated the atoms as two-level systems with a circular transition. Similar results can however also be obtained using instead atoms with a single ground state $\ket{g}_{\boldsymbol j}$ and three excited states $\ket{e_i}_{\boldsymbol j}$ ($i=x,y,z$), where $i$ denotes the dipole orientation axis. The non-hermitian dipole-dipole interaction Hamiltonian from Eq.~\eqref{mathcalH} generalizes to 
\begin{equation*}
H_\text{dip}=-i(\gamma_e/2)\sum_{{\boldsymbol j},{\boldsymbol j}'}\sum_{i,i'}\hat {\boldsymbol G}_{i,i'}({{\boldsymbol r}_{\boldsymbol j}-{\boldsymbol r}_{\boldsymbol j'}})\sigma_{\boldsymbol j,i}^+\sigma_{{\boldsymbol j}',i'}^-,
\end{equation*}
where $\sigma_{\boldsymbol j,i}^-=\ket{g}_{\boldsymbol j}\!\bra{e_i}$, which now mixes states with different polarizations.
Diagonalizing this Hamiltonian, we obtain a degenerate pair of dark and bright states, polarized in the $x-y$ plane, with decay rates represented in Fig.~\ref{fig:FigIso}. Notably, these decay rates remain close to the values obtained for two-level atoms. 

\end{document}